\documentclass[aps,prd,longbibliography,nofootinbib]{revtex4-2}
\usepackage{graphicx}% Include figure files
\usepackage{dcolumn}% Align table columns on decimal point
\usepackage{amsmath,amssymb,epsfig}
\usepackage{paralist}
\usepackage{comment}
\usepackage{multirow}
\usepackage{color,soul}
\usepackage{suffix}
\usepackage{mathtools}

\usepackage{siunitx}
\usepackage{longtable}
\usepackage{array}
\usepackage{natbib}
\usepackage{hyperref}
\hypersetup{colorlinks=true,allcolors=blue}

\usepackage{booktabs}
\usepackage{tikz}
\usepackage{pgfplots}
\pgfplotsset{compat=1.18}

\usepackage{ragged2e}

\begin{document}

\title{Lunar Dust: Formation, Microphysics, and Transport}

\author{Slava G. Turyshev}   

\affiliation{ 
Jet Propulsion Laboratory, California Institute of Technology,\\
4800 Oak Grove Drive, Pasadena, CA 91109-0899, USA
}%

\date{\today}% It is always \today, today,

\begin{abstract}

Lunar dust---the sub-millimeter fraction of the regolith---controls the optical, thermophysical, electrical, mechanical, and environmental behavior of the Moon's surface. These properties set the performance envelopes of remote-sensing retrievals, regolith geotechnics, volatile cycles, and exploration systems, while also posing operational and biomedical risks. We synthesize Apollo sample analyses and in-situ observations (Surveyor, Lunokhod, Apollo) with contemporary datasets from the LRO Diviner Lunar Radiometer, the LADEE/LDEX exospheric dust measurements, and Chang'e-4 Lunar Penetrating Radar (LPR). We also incorporate 2024-2025 results: Chandrayaan-3 ChaSTE thermophysics at the Vikram lander's site, SCALPSS plume-surface diagnostics from Intuitive Machines Mission 1 (IM-1), and Negative Ions at the Lunar Surface (NILS) detections of a dayside near-surface H$^{-}$ population on Chang'e-6. The review links (i) production and modification processes to (ii) grain-scale physical/chemical/electrical/optical/mechanical properties, then to (iii) mobilization pathways (meteoroid ejecta, electrostatic hopping, rocket-plume entrainment), and finally to (iv) region-specific design ranges across maria, highlands, pyroclastic units, magnetic swirls, and  permanently shadowed regions (PSRs). We quantify temperature-illumination dependence across day/night and PSR-equator regimes through a two-channel $k(T,\rho)$ model and charge-relaxation scaling. We provide closed-form expressions for adhesion-aware lift thresholds and for near-surface (0--3 m) dust transport at apex/hover heights as functions of sheath structure. The result is a design-ready set of relations, figures, and tables that propagate microphysics and composition into engineering parameters for upcoming landed and rover operations.
    
\end{abstract}

\maketitle

\tableofcontents

\section{Introduction}

Airless-body regolith evolves under micrometeoroid bombardment, solar-wind implantation, and thermal cycling. That evolution governs the microstructure and composition of the sub-millimeter fraction (``lunar dust''), which, in turn, sets thermal conductivity $k$, real permittivity $\varepsilon'$, loss tangent $\tan\delta$, friction angle $\varphi$, cohesion $c$, optical maturity, and charge exchange with the near-surface plasma. 
Microstructure and composition in turn determine thermophysical, electrical, optical, and mechanical properties that control dust behavior under changing illumination and plasma conditions. These properties determine when dust is liberated, where it accumulates, and how it ages. These same properties govern operational risks (e.g., fouling, abrasion, charging) and scientific observables (e.g., reflectance, Christiansen Feature (CF) position, dielectric). Our intent is to connect the chain from production mechanisms to emergent properties to dynamical consequences and, finally, to engineering decisions.

The exploration-relevant uncertainties today are set by three recent developments:\footnote{{\it Abbreviations used:}
Apollo sample program (Apollo); Commercial Lunar Payload Services (CLPS); Chandra's Surface Thermophysical Experiment (ChaSTE); Christiansen Feature (CF); Diviner Lunar Radiometer (Diviner); gravitational acceleration on the Moon ($g_{\rm Moon}$); Intuitive Machines Mission 1 (IM-1); Lunar Atmosphere and Dust Environment Explorer (LADEE); Lunar Dust Experiment on LADEE (LDEX); Lunar Penetrating Radar (LPR); Moon Mineralogy Mapper on Chandrayaan-1 (M$^{3}$); nanophase metallic iron (npFe$^{0}$); Negative Ions at the Lunar Surface (NILS); permanently shadowed region (PSR); particle-size distribution (PSD); plume--surface interaction (PSI); specific surface area (SSA).} (1) cm-scale thermophysical structure resolved by Diviner inversions and now anchored in situ by ChaSTE at high southern latitude; (2) direct near-surface constraints on sheath composition from NILS, which require negative ions to be included in daytime charging budgets; and (3) a new generation of plume-surface diagnostics (SCALPSS-class stereo photogrammetry plus ejecta counters) necessary for bounding erosion rates, particle-size distributions, and angular flux during landings. Together these advances enable engineering-grade ranges that were not available at the time of earlier reviews.

This review is motivated by two gaps. First, new landed/orbital constraints (ChaSTE, NILS, SCALPSS) have not been fused with Apollo--Diviner--LADEE--LPR into a quantitatively self-consistent reference that spans depth, latitude, and terrain. Second, the literature connecting grain-scale properties (particle-size distribution, angularity, agglutinate and nanophase metallic iron content, composition) to design-relevant envelopes for mobilization pathways (meteoroid ejecta, electrostatic lofting, rocket-plume entrainment) is scattered. Here we close both gaps by (a) compiling region-specific ranges with uncertainty where available, and (b) providing equations with labeled parameters and default reference values, propagating microstructure to bulk properties, and integrating those into mobilization budgets.

We (a) unify production and modification mechanisms into three microphysical levers that dominate downstream behavior (particle-size distribution and angularity, agglutinate fraction with nanophase Fe$^{0}$, and composition/oxidation state); (b) quantify how those levers map to specific surface area, vertical density structure, permittivity, thermal conductivity, and optical maturity; (c) derive lift thresholds including cohesion and show where pre-liberation is required; (d) integrate meteoroid ejecta, electrostatic mobilization, and rocket-plume entrainment across regions; and (e) tabulate region-specific reference values for use in design studies at the poles, over swirls, and at landing sites.

Our objectives are (i) to compile production and modification mechanisms with first-order scalings; (ii) to provide ranges and recommended reference values for physical, chemical, dielectric, thermal, mechanical, and optical parameters by region; (iii) to derive compact bridging relations that connect microstructure to $k(T,\rho)$, $\varepsilon'(\rho,\mathrm{composition})$, and electrostatic lift with adhesion; and (iv) identify colocated measurements that most efficiently reduce design margins at the poles, magnetic swirls, pyroclastics, and landing sites.

To reach our objectives, we draw on Apollo geotechnics and petrography \citep{Carrier1991,Carrier2003}, Diviner thermophysics \citep{Bandfield2011,Hayne2017,Paige2010}, {LADEE}/{LDEX} exospheric dust measurements \citep{Horanyi2015}, Chang'e-4 {LPR} dielectric profiles \citep{Dong2021,Giannakis2021}, syntheses of space-weathering and maturity \citep{Morris1978,Lucey2006,Lucey2017CF,Taylor2001,Wang2017}, tribo/photo-charging \citep{Colwell2007,Stubbs2006,Forward2009,Sternovsky2002}, composition and mineralogy \citep{Papike1998,Green2011,Taylor2010}, pyroclastics \citep{Allen2012,Delano1986,Hui2018}, plume--soil interaction \citep{Carroll1971,Metzger2011,Metzger2024}, and health hazards \citep{Gaier2005,Gaier2011,Linnarsson2012,Lam2013,Pohlen2022,Wallace2009react}.
Additional 2024--2025 results include Chandrayaan-3 ChaSTE in-situ thermophysical measurements at the Vikram lander's site (69.373${}^\circ$~S,\,32.319${}^\circ$~E) (thermal conductivity at 80 mm depth and local packing density) \cite{Mathew2025ChaSTE}, a consolidated review of plume--surface interaction experiments and SCALPSS campaign status for Intuitive Machines Mission 1 (IM-1) and subsequent landings \cite{Cuesta2025PSI}, and Chang'e-6 Negative Ions at the Lunar Surface (NILS) detections of a dayside near-surface H$^{-}$ layer \cite{CanuBlot2025NILS,Wieser2025NILSHminus}; we also include LADEE/LDEX meteoroid-stream responses that quantify event-driven dust variability \cite{Szalay2018Geminids}.

New constraints derived here will make an engineering--grade synthesis feasible:\footnote{Programmatically, recent dust-mitigation roadmaps and flight demonstrations provide additional quantitative data for subsystem requirements. The NASA Dust Mitigation Technology Roadmap \cite{Fritz2024Roadmap} organizes passive/active/dust-tolerant strategies and associated standards for test and verification, and an electrodynamic dust shield (EDS) has now been demonstrated on the lunar surface with reported dust-removal fractions of 82\% (radiator) and 97\% (glass) over several cleaning cycles \cite{Fritz2024Roadmap,Buhler2025EDS,Abel2023DustGuide}.
} (i) centimeter--depth thermophysics anchored \emph{in situ} by ChaSTE at high southern latitude; (ii) daytime near--surface negative ions (H$^-$) that modify sheath composition and charging; and (iii) plume--surface diagnostics that bound erosion, PSD, and angular flux. We intend to  fuse these with Apollo--Diviner--LADEE--LPR to produce a single, quantitative reference spanning depth, latitude, and terrain. Prior reviews treat subsets of this chain (thermophysics, dust charging, or PSI) in isolation; here we provide a single, internally consistent map from sources $\rightarrow$ properties $\rightarrow$ transport $\rightarrow$ subsystem impacts, with labeled equations and region--specific priors to enable immediate design use. Throughout we treat the values compiled in Tables~\ref{tab:regional} and \ref{tab:properties} as engineering priors with explicit provenance (typically bracketing inter-site variability and inversion/model systematics), not as specifications; wherever possible, designers should replace these priors with site- and hardware-specific measurements.

This paper is organized as follows. Section~\ref{sec:formation} synthesizes production and modification pathways and provides first-order scalings. Section~\ref{sec:phys-chem} compiles grain-scale properties (PSD, angularity/SSA, vertical density, and composition) that control contacts and observables. Sections~\ref{sec:mech-prop} and \ref{sec:thermphys-prop} connect that microstructure to geotechnical and thermophysical behavior via Mohr-Coulomb shear parameters, adhesion at asperities, and a two-channel $k(T,\rho)$ model. Section~\ref{sec:elec-dielec-prop} develops dielectric mixing, charge relaxation, and sheath-field budgets, culminating in an adhesion-aware lift criterion and its scaling. Section~\ref{sec:opt-prop} summarizes optical and maturity diagnostics relevant to dust coatings and sorting. Section~\ref{sec:mobilization-transport} integrates the mobilization pathways (meteoroid ejecta, electrostatic hopping in the first meters, and rocket--plume entrainment); it also quantifies near-surface (0--3~m) trajectories/hover heights versus sheath structure. Sections~\ref{sec:reg_refvalues} and \ref{sec:eng_refvalues} provide region-specific priors and global engineering ranges. Section~\ref{sec:design} discusses model relations useful in design; it also provides measurement priorities that most efficiently reduce design margins for upcoming polar and mid-latitude operations. In Section~\ref{sec:concl} we conclude and summarize our findings.

\section{Formation and modification mechanisms}
\label{sec:formation}

The production and modification pathways establish the levers that control dust behavior: particle-size distribution (PSD) and angularity, agglutinate fraction with npFe$^{0}$, and composition/oxidation state (see Table~\ref{tab:nomenclature} for notations). We quantify each process and provide scalings that are used downstream in Secs.~\ref{sec:mech-prop}--\ref{sec:elec-dielec-prop}.

\subsection{Impact comminution and agglutination}
\label{sec:impacts}

Micrometeoroid impacts fragment rock and soil and produce glass-welded agglutinates embedding npFe$^{0}$, which dominate mature-soil optical behavior \cite{Carrier1991,Taylor2001}. The $< 1$ mm soil fraction is well described by a lognormal PSD, $\ln D \sim \mathcal{N}(\mu,\sigma^{2})$, with median $D_{50}\approx 40\text{--}130~\mu$m and $\sigma \approx 0.8\text{--}1.2$ by count \cite{Carrier2003}. Agglutinate fraction and $I_{\rm s}$/FeO increase with exposure \citep{Morris1978}.

\begin{table}[t]
\caption{Symbols used repeatedly throughout. Also, we use $g_{\mathrm{Moon}}=1.62\,\mathrm{m\,s^{-2}}$; 
$a$ is grain radius.}
\label{tab:nomenclature}
\centering
\begin{tabular}{@{}ll@{}}
\hline
Symbol & Definition [units]\\
\hline
$\rho,\rho_{s}$ & bulk, solid density [kg\,m$^{-3}$]\\
$n$ & porosity $=1-\rho/\rho_{s}$ [--]\\
$k,\ell,A$ & thermal conductivity parameters in Eq.~(\ref{eq:k-twochannel});
$k$~[\si{\watt\per\meter\per\kelvin}], $\ell$~[--],
$A$~[\si{\watt\per\meter\per\kelvin\tothe{4}}]\\
$E,\lambda$ & near-surface electric field amplitude, e-fold length
[\si{\volt\per\meter}, \si{\meter}]\\
$\Delta V_{\rm sh}$ & near-surface sheath potential drop $\int_{0}^{\infty}\!E(z)\,dz\simeq E_{0}\,\lambda$ [\si{\volt}]\\
$\phi_g$ & grain potential [\si{\volt}]\\
$W, R$ & work of adhesion, asperity radius
[\si{\joule\per\meter\squared}, \si{\meter}]\\
$\Gamma$ & thermal inertia $\sqrt{k\,\rho\,c_p}$ [J\,m$^{-2}$\,K$^{-1}$\,s$^{-1/2}$]\\
$c_p$ & specific heat capacity  [J\,kg${}^{-1}$\,K${}^{-1}$]\\
$\varepsilon'$, $\tan\delta$ & real permittivity, loss tangent [--]\\
$\phi, c$ & friction angle, cohesion [\si{\degree}, \si{\pascal}]\\
\hline
\end{tabular}
\end{table}

Measured Brunauer--Emmett--Teller specific surface area (SSA) spans 0.02--0.78 m$^{2}$ g$^{-1}$, with a widely used reference near 0.5 m$^{2}$ g$^{-1}$ (spherical-equivalent diameter 3--4 $\mu$m). A compact vertical density parameterization that matches Apollo cores and Diviner inversions is
{}
\begin{equation}
\label{eq:rhoz}
\rho(z)=\rho_{\infty}-\bigl(\rho_{\infty}-\rho_{0}\bigr)\,e^{-z/z_{c}},
\end{equation}
with reference values $\rho_{0}\approx 1.4\,\mathrm{g\,cm^{-3}}$ at the surface, $\rho_{\infty}\approx 1.85\,\mathrm{g\,cm^{-3}}$ at depth, and $z_{c}=0.15$--$0.30\,\mathrm{m}$.

\subsection{Space weathering}

Solar wind and micro-impacts generate nanophase metallic iron (npFe$^{0}$) in 10--200 nm rims and within agglutinitic glass, lowering albedo, steepening spectral slopes, and suppressing mafic band depths \citep{Lucey2006,Taylor2001}. A first-order attenuation of a band depth (BD) with cumulative solar-wind exposure is
\begin{equation}
\label{eq:BD}
{\rm BD} \approx {\rm BD}_{0} \exp(-\alpha\,\phi_{\mathrm{sw}}), \qquad 0.1 \lesssim \alpha \lesssim 0.5,
\end{equation}
where $\phi_{\rm sw}$ is a dimensionless solar-wind ``dose'' variable proportional to the cumulative solar-wind fluence at the surface. We write
\begin{equation}
\phi_{\rm sw}(t) \equiv \frac{\Phi_{\rm sw}(t)}{\Phi_{\rm ref}}, \qquad 
\Phi_{\rm sw}(t) = \int_{0}^{t} F_{\rm sw}(t')\,{\rm d}t',
\end{equation}
where $F_{\rm sw}$ is the solar-wind ion flux (dominated by protons) in particles cm$^{-2}$\,s$^{-1}$, and $\Phi_{\rm sw}$ is the cumulative proton fluence in particles cm$^{-2}$. $\Phi_{\rm ref}$ is a characteristic fluence scale, chosen here as
$\Phi_{\rm ref} \sim 10^{23}\ {\rm cm^{-2}}$, comparable to the space-weathering doses inferred for mature lunar soils at 1~AU. For design use, a representative proton flux $F_{\rm sw} \simeq 10^{8}\ {\rm cm^{-2}\,s^{-1}}$ gives
$\Phi_{\rm sw} \sim 3\times 10^{22}$--$3\times 10^{23}\ {\rm cm^{-2}}$ over $10^{7}$--$10^{8}$~yr of exposure, corresponding to $\phi_{\rm sw} \sim 0.3$--$3$. In this parametrization both $\alpha$ and $\phi_{\rm sw}$ are dimensionless, and the product $\alpha \phi_{\rm sw}$ is a dimensionless space-weathering ``dose'' variable for use in photometric/maturity correction. The Diviner CF shifts with both composition and weathering state \citep{Lucey2017CF,Wang2017} and is used in Section~\ref{sec:elec-dielec-prop} to cross-calibrate composition versus maturity.

\subsection{Thermal fatigue}
\label{sec:thermal-fatigue}

Thermoelastic stresses scale as $\sigma_{\rm th} \sim E\alpha_{T}\Delta T/(1-\nu)$. With $E=50$--100 GPa, $\alpha_{T}=5$--10$\times 10^{-6}$ K$^{-1}$, $\Delta T=200$--300 K, $\nu\approx 0.25$, stresses of order 10--100 MPa arise, consistent with laboratory cycling that produces crack advance and fines \citep{Patzek2022}. Diviner rock-abundance decay vs. age indicates ongoing breakdown on Myr--100 Myr timescales \citep{Bandfield2011}.

\subsection{Volcanic pyroclastics}
Fire-fountain eruptions produced glass beads and shards (Apollo 15 green; Apollo 17 orange/black). Representative compositions: FeO 16--25 wt\%, TiO$_2$ 9--16 wt\% for high-Ti glasses \citep{Delano1986,Allen2012,Hui2018}. These glasses generate fines with distinct mid-IR emissivity and CF positions.

\emph{From sources to properties.} The production pathways discussed above in  Sec.~\ref{sec:formation} set three levers that dominate the downstream physics: (i) particle-size distribution and angularity (impact comminution, thermal fatigue), (ii) glassy agglutinate content with npFe$^{0}$ (space weathering), and (iii) composition and oxidation state (basaltic versus anorthositic, pyroclastic glasses). Secs.~\ref{sec:phys-chem}--\ref{sec:thermphys-prop} quantify how these levers map to specific surface area, vertical density structure, dielectric response, and spectral behavior that the rest of the paper builds upon.

\section{Physical and chemical properties}
\label{sec:phys-chem}

Given the pathways in Sec.~\ref{sec:formation}, we compile grain-scale properties that set contact mechanics, transport, and remote-sensing response: PSD, SSA, vertical density structure, and bulk chemistry/mineralogy.

\subsection{Particle-size distribution (PSD), shape, surface area}

Mature soils exhibit a broad, well-graded PSD with median diameter $D_{50}\approx 40$--$130~\mu$m and abundant sub-$20~\mu$m fines by count \citep{Carrier2003}. Irregular, angular grains raise SSA and promote interlocking at contacts. Measured Brunauer--Emmett--Teller (BET) SSA spans $0.02$--$0.78$ m$^{2}$ g$^{-1}$, with a widely used reference value near $0.5$ m$^{2}$ g$^{-1}$ (spherical-equivalent diameter $3$--$4~\mu$m) \cite{Carrier1991,Carrier2003,Cadenhead1977}. 

\subsection{Density, porosity, and vertical structure}
\label{sec:den-poro}

A compact representation of bulk density $\rho(z)$ from (\ref{eq:rhoz}) is
\[
\rho(z) = \rho_{\infty} - (\rho_{\infty}-\rho_{0})\,e^{-z/z_{c}},
\]
with $\rho_{0}=1.4$ g cm$^{-3}$, $\rho_{\infty}=1.85$ g cm$^{-3}$, $z_{c}=0.15$--0.30 m. Near-surface porosity is typically 40--50 percent. Regolith thickness: maria 3--5 m; highlands 10--20 m \citep{Carrier1991}. These structures are consistent with Diviner vertical $k$ inversions \citep{Hayne2017}.

\subsection{Mineralogy and chemistry}
\label{sec:mineralogy}

Maria are basaltic (FeO 15--22 wt\%, TiO$_2$ 1--13 wt\%); highlands are anorthositic (plagioclase rich, low FeO) \citep{Papike1998,Lucey2006,Taylor2010}. Space weathering adds npFe$^{0}$ and amorphous rims that dominate optical maturity metrics.

From microstructure to mechanics. Given the PSD, angularity, and SSA, the Mohr--Coulomb parameters and contact physics follow: higher angularity and packing density raise $\phi$ and elevate $c$ through interlocking; increased fines raise adhesion and increase the work required to detach grains from contacts in vacuum. These relations motivate the geotechnical formulations collected in Sec.~\ref{sec:mech-prop}.

\section{Mechanical properties and contact physics}
\label{sec:mech-prop}

\subsection{Shear strength}
\label{sec:shear}

Under drained, vacuum conditions the peak shear strength of the lunar regolith on a plane at effective normal stress $\sigma$ is described by the Mohr--Coulomb criterion
\begin{equation}
\label{eq:morn-coul}
\tau(\sigma,z) \;=\; c(z) \;+\; \sigma \,\tan\phi(z),
\end{equation}
where $\tau$ is shear stress at failure [Pa], $c$ is the apparent cohesion [Pa] arising from interlocking, adhesion and any vacuum cementation,  $\sigma$ is the \emph{effective} normal stress on the failure plane [Pa] (pore pressure $u=0$ in vacuum, so $\sigma=\sigma_n$), and $\phi$ is the friction angle [deg], which increases with relative density and confinement.
Apollo/Lunokhod and analog testing indicate 
$\phi = 30^\circ$--$50^\circ$ at the surface, approaching $\sim 55^\circ$ with depth/relative density, and 
$c = 0.1$--$1$~kPa at the surface, increasing to $>3$~kPa by 0.5--1~m depth\footnote{Ranges compile classic and modern syntheses cited herein.}  \cite{Carrier1991,Connolly2023}.
These trends are consistent with angular grains and broad PSDs that raise interlocking and contact density with depth.

To relate strength to depth, we again use the bulk-density profile of Eq.~(\ref{eq:rhoz}), which gives the vertical overburden (effective) stress
\begin{equation}
\label{eq:eff-stress}
\sigma_v(z)=\int_0^{z}\rho(\zeta)\,g_{\rm Moon}\,d\zeta
= g_{\rm Moon}\!\left[\rho_{\infty}z-(\rho_{\infty}-\rho_{0})z_{c}\left(1-e^{-z/z_{c}}\right)\right].
\end{equation}

\paragraph{PSR frost-cement contingency.}
If a volumetric ice fraction $f_{\rm ice}$ is present within the near-surface pore space (PSR floors/rims), apparent cohesion and (possibly) friction angle are modified. For design, carry a parametric contingency:
\begin{equation}
c_{\rm PSR}(z) \equiv c(z) + \Delta c_{\rm ice},\qquad
\phi_{\rm PSR}(z) \equiv \phi(z) + \Delta\phi_{\rm ice},
\label{eq:PSR-eq_6}
\end{equation}
with increments scaled to $f_{\rm ice}$,
\begin{equation}
\Delta c_{\rm ice} = \alpha_c\,\frac{f_{\rm ice}}{1-n(z)}\;\sigma_{\rm ice}(T),\qquad
\Delta\phi_{\rm ice} = \alpha_\phi\,f_{\rm ice},
\label{eq:PSR-eq_7}
\end{equation}
where $n(z)$ is porosity from Eq.~(\ref{eq:rhoz}), $\sigma_{\rm ice}(T)$ is the unconfined compressive strength of ice at temperature $T$, and $\alpha_c,\alpha_\phi$ are dimensionless design coefficients to be bracketed in sensitivity studies (e.g., $\alpha_c\in[0.1,1]$, $\alpha_\phi\in[0^\circ,4^\circ]$). Use local $T$ from Sec.~\ref{sec:thermphys-prop} and update trafficability envelopes in Sec.~\ref{sec:trafficability} and Table~\ref{tab:design-mapping} accordingly.

As an illustrative design calculation (maria prior: $\rho_{0}=1.4~\mathrm{g\,cm^{-3}}$, $\rho_{\infty}=1.85~\mathrm{g\,cm^{-3}}$, $z_{c}=0.20$~m), at $z=0.30$~m the overburden is $\sigma_v\simeq7.9\times10^{2}$~Pa. With $\phi=40^\circ$ and $c=0.5$~kPa this yields
\[
\tau \approx c + \sigma_v \tan\phi \;\simeq\; 0.5~\mathrm{kPa} + (0.79~\mathrm{kPa})\tan 40^\circ \;\simeq\; 1.16~\mathrm{kPa},
\]
consistent with Apollo geotechnical ranges and the increase of $(\phi,c)$ with density at depth \cite{Carrier1991,Connolly2023}.

\subsection{Adhesion and detachment thresholds}

At a single asperity of radius $R$,  the Johnson--Kendall--Roberts (JKR) estimate \cite{Johnson1971JKR} for pull-off force for a sphere on a flat is
\begin{equation}
F_{\mathrm{adh}} \approx \tfrac{3}{2}\pi W R.
\label{eq:JKR-pull-off}
\end{equation}
With $W \approx 0.05$--$0.2$~J\,m$^{-2}$ and $R \approx 1$--$10~\mu$m, (\ref{eq:JKR-pull-off}) results in $F_{\mathrm{adh}}\approx 0.24$--$9.42~\mu$N for the stated ranges (for $W=0.1$~J\,m$^{-2}$, $R=5~\mu$m, $F_{\mathrm{adh}}=2.36~\mu$N). For a 1 $\mu$m grain with density $\rho_{\mathrm{p}}=3100$~kg~m$^{-3}$, the gravitational force $m g_{\mathrm{Moon}} \approx 2.10\times 10^{-14}$ N (eight orders of magnitude below the adhesion force for the parameters used here). Consequently, pre-liberation by vibration, micro-impacts, or gas shear is typically required before electrostatic acceleration can act \cite{Walton2007,Berkebile2012}; Sec.~\ref{sec:charg-sheath-lift} quantifies the field strengths needed when $F_{\mathrm{adh}}$ is included.

\paragraph*{JKR vs.\ DMT regime:}
 In the Derjaguin-Muller-Toporov (DMT) limit \cite{Derjaguin1975_DMT,Maugis1992_JKR_DMT} appropriate to stiffer materials and smaller contact radii, the pull-off force is $F_{\rm adh}^{\rm DMT} = 2\pi W R$ (compare to (\ref{eq:JKR-pull-off})).
The appropriate contact model can be gauged by the Tabor parameter $\mu = ({R W^{2}}/{E^{*2} z_{0}^{3}})^{1/3}$,
with reduced modulus $E^*$ and intermolecular cutoff $z_{0}\sim 0.3$--$0.4$~nm. For lunar-like values $R \sim 1$--$10\,\mu$m, $W \sim 0.1$\,J\,m$^{-2}$, and $E^*\!\sim\!50$--$100$\,GPa, $\mu \gtrsim 1$--$10$, placing contacts in the JKR regime. If $\mu \lesssim 1 $, the DMT limit applies, with $F_{\rm adh} \approx 2\pi W R$. Given that $F_{\rm adh}\gg mg_{\rm Moon}$ across the micron range, both models lead to the same design implication: pre-liberation (vibration, micro-impacts, or gas shear) is required before sheath acceleration is effective. We adopt JKR here and propagate $(W,R)$ explicitly into the lift requirement via Eq.~(\ref{eq:E-adh}). 

\subsection{Trafficability and wheel--soil interaction (design bridge)}
\label{sec:trafficability}

For a driven wheel (radius $R_{\rm w}$, width $b$) carrying normal load $W$, the contact normal stress is 
$\sigma \simeq W/(b\,\ell_c)$ with contact length $\ell_c \sim \sqrt{2R_{\rm w} z}$ for sinkage $z \ll R_{\rm w}$. 
The available shear stress at the interface follows Mohr--Coulomb criterion (see Sec.~\ref{sec:shear}): 
\begin{equation}
\tau_{\max} = c + \sigma \tan\phi,
\label{eq:tau_max_mc}
\end{equation}
with cohesion $c$ and friction angle $\phi$ taken from Sec.~\ref{sec:mech-prop} and Table~\ref{tab:regional}. 
A first--order upper bound on tractive force for one wheel is then
\begin{equation}
T_{\max} \simeq \tau_{\max}\,b\,\ell_c, 
\qquad 
\Rightarrow\quad 
T_{\max} \simeq \Big(c + \frac{W}{b\,\ell_c}\tan\phi \Big)\,b\,\ell_c,
\label{eq:tractive_bound}
\end{equation}
which elevates with increasing $\phi$ and $c$ (depth/relative density), consistent with Apollo geotechnics and modern syntheses (Sec.~\ref{sec:mech-prop}). 
The corresponding rolling resistance force can be estimated as 
\begin{equation}
F_{\rm rr} \simeq k_{\rm rr}\,W, 
\qquad k_{\rm rr}\approx \frac{z}{R_{\rm w}},
\label{eq:rolling_res}
\end{equation}
so denser near--surface structure Eq.~(\ref{eq:rhoz}) and higher $\phi$ reduce sinkage $z$ and $k_{\rm rr}$. 

\paragraph{Pressure--sinkage scaling.}
For pressure $p$ under a plate of width $b$, the classical pressure--sinkage form 
\begin{equation}
p(z) \approx \Big(\frac{k_{c}}{b} + k_{\phi}\Big) z^{\,n}
\label{eq:bekker}
\end{equation}
is a convenient parameterization; $(k_c,k_\phi,n)$ are site--specific and should be calibrated with preflight terrain analogs using the density profile of Eq.~(\ref{eq:rhoz}).\footnote{Use $\phi$ and $c$ from Sec.~\ref{sec:mech-prop}; initialize $(k_c,k_\phi,n)$ from analog tests, then adjust to match in situ slip--sinkage telemetry.} 
Combining Eqs.~(\ref{eq:tau_max_mc})--(\ref{eq:bekker}) with regional priors in Table~\ref{tab:regional} gives a compact mobility envelope \emph{without} invoking additional constitutive assumptions.

\paragraph{Design use.}
Given $W$, $R_{\rm w}$, $b$, and site priors $(\rho(z),\phi,c)$, Eqs.~(\ref{eq:tractive_bound})--(\ref{eq:rolling_res}) bound gradeability and power at slip. 
PSR rims and cold slopes tend to exhibit higher relative density near the surface (Sec.~\ref{sec:thermphys-prop}), hence reduced $z$ and $k_{\rm rr}$; pyroclastics and immature swirl lanes can deviate systematically (Tables~\ref{tab:regional}, \ref{tab:properties}). 

\section{Thermophysical properties}
\label{sec:thermphys-prop}

Thermophysics mediates between structure and environment. The effective thermal conductivity of lunar regolith (or dust), $k(T,\rho)$,  is a function of temperature $T$ and bulk density $\rho$. Depth-dependent $k(T,\rho)$ determines the diurnal temperature wave and, through it, photoemission and charge relaxation times relevant to Section~\ref{sec:elec-dielec-prop}.

\subsection{Conductivity and thermal inertia}
\label{sec:cond-thinert}

A contact-plus-radiative representation captures both solid-network and pore-radiative pathways:
\begin{equation}
k(T,\rho) = k_{\mathrm{cond}}(\rho) + A\,T^{3}, \qquad
k_{\mathrm{cond}}(\rho) = k_{\mathrm{cond},0}\Big(\frac{\rho}{\rho_{s}}\Big)^{\ell},
\label{eq:k-twochannel}
\end{equation}
with $\ell \approx 2$ to $3$, solid density $\rho_{s}$, and radiative coefficient $A$ [W\,m$^{-1}$\,K$^{-4}$] that scales with porosity and pore-size statistics. The radiative coefficient $A$ is quantified in Sec.~\ref{sec:rad-term}. Diviner retrievals constrain $k$ near the surface to $\sim 7.4 \times 10^{-4}\ \mathrm{W\,m^{-1}\,K^{-1}}$, increasing to $\sim 3.4 \times 10^{-3}\ \mathrm{W\,m^{-1}\,K^{-1}}$ by $\sim 1\ \mathrm{m}$ depth, while ChaSTE measured $k \approx (1.15$ to $1.24)\times 10^{-2}\ \mathrm{W\,m^{-1}\,K^{-1}}$ at $z=0.08\ \mathrm{m}$ at 69.373${}^\circ$~S, consistent with local densification and the depth dependence in (\ref{eq:k-twochannel}). Thermal inertia $\Gamma=\sqrt{k\,\rho\,c_p}$, with  $c_p$ being specific heat capacity  in  [J\,kg${}^{-1}$\,K${}^{-1}$], thermal inertia $\Gamma$ is typically $40$--$70$ J m$^{-2}$\,K$^{-1}$\,s$^{-1/2}$ for fines, increasing with rock abundance and latitude \citep{Bandfield2011}. PSRs reach 25--40 K \citep{Paige2010}. 

\begin{figure}[h]
\centering
\begin{tikzpicture}
\begin{axis}[
  width=0.62\linewidth,
  height=0.41\linewidth,
  xlabel={Depth $z$ (m)},
  ylabel={$k(z)$ (W m$^{-1}$ K$^{-1}$)},
  xmin=0, xmax=1.0,
  ymin=5e-4, ymax=2e-2,
  ymode=log, grid=both,
  tick label style={/pgf/number format/fixed},
]
% simple exponential increase: k(z) = k_inf - (k_inf - k0) exp(-z/zc)
\pgfmathsetmacro{\kzero}{7.4e-4}
\pgfmathsetmacro{\kinf}{3.4e-3}
\pgfmathsetmacro{\zc}{0.20}
\addplot+[domain=0:1,samples=100, mark=*, mark size=1.1pt,
          mark options={line width=0.2pt}]
  ({x},{ \kinf - (\kinf - \kzero) * exp(-x/\zc) });
% ChaSTE point at 0.08 m, k ~ 1.2e-2
\addplot+[only marks,mark=*, mark size=1.4pt] coordinates {(0.08,0.012)};
\end{axis}
\end{tikzpicture}
\caption{Illustrative conductivity--depth profile $k(z)$ anchored to Diviner day/night trends  and the ChaSTE in-situ point at $z\approx 0.08$ m ($k\simeq\SI{1.2e-2}{\watt\per\meter\per\kelvin}$) \cite{Hayne2017,Paige2010,Mathew2025ChaSTE} closes the long--standing gap between Diviner surface inversions and near--subsurface $k(z)$. The elevated $k$ at 8 cm is consistent with higher density and local disturbance from landing, emphasizing centimeter-scale structure in $k(T,\rho)$. }
\label{fig:kdepth}
\end{figure}
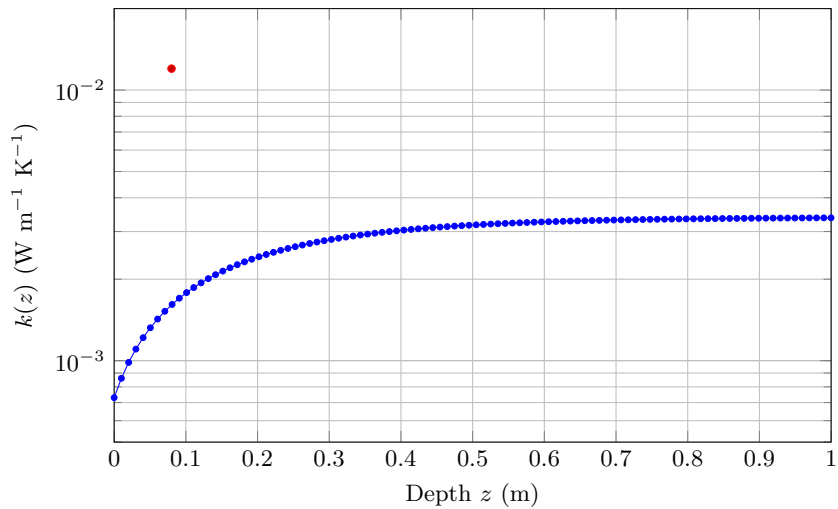

New in-situ measurements by the Chandrayaan-3 ChaSTE probe at the Vikram site (69.373${}^\circ$~S,\,32.319${}^\circ$~E) report $k=(1.15$--$1.24)\times 10^{-2}$ W m$^{-1}$ K$^{-1}$ at a depth of 80 mm and indicate a local packing density $\rho \approx 1.94\times 10^{3}$~kg~m$^{-3}$ \cite{Mathew2025ChaSTE}. The relatively high $k$ compared to near-surface Diviner inversions is consistent with (i) density increase with depth, (ii) local compaction and fines redistribution associated with touchdown, and (iii) the explicit $k(T,\rho)$ dependence in the two-channel (contact + radiative) model used here. These ChaSTE values provide an anchor for calibrating $k(T,\rho)$ at centimeter depths in polar design studies \cite{Mathew2025ChaSTE}. See Fig.~\ref{fig:kdepth} for an illustrative $k(z)$ profile anchored by Diviner near the surface and by the ChaSTE point at $z \approx 0.08$ m.

Coupling to the electrical environment. The $k(T,\rho)$ structure regulates the near-surface diurnal temperature wave and hence photoemission, conductivity of the pore network, and the time constants of charge relaxation. At high latitudes and in PSRs, low $T$ and high porosity lengthen charge relaxation times, biasing the electrical regime toward persistent charge patches and large $|E|$ at boundaries. These couplings are taken up in Sec.~\ref{sec:elec-dielec-prop}.

\paragraph*{Recommended parameter ranges and sensitivities (design use).}
To make Eq.~(\ref{eq:k-twochannel}) directly actionable in trades, we recommend carrying explicit ranges and first-order sensitivities:
\begin{itemize}
  \item Exponent: $\ell = 2.0\text{--}3.0$ (dimensionless). \emph{Sensitivity:}
  {}
\begin{equation}
\label{eq:k-rho-sensit}
  \frac{\partial \ln k}{\partial \ln \rho} \;=\; \frac{k_{\rm cond}}{k}\,\ell
  \quad\text{with}\quad
  k_{\rm cond} = k_0\Big(\frac{\rho}{\rho_s}\Big)^{\ell}.
\end{equation}
  \item Radiative coefficient: $A = (3.6\times10^{-12}\text{--}3.6\times10^{-11})\ {\rm W\,m^{-1}\,K^{-4}}$
  (consistent with the day/night asymmetry used later in Sec.~\ref{sec:rad-term}).
  \emph{Sensitivity:}
\begin{equation}
\label{eq:k-T-sensit}
  \frac{\partial \ln k}{\partial \ln T} \;=\; \frac{3A T^{3}}{k}\,.
\end{equation}
  \item Contact term scale $k_0$: set by calibration to the Diviner and ChaSTE anchors already cited:
\begin{align}
  k(z\to 0) &\,\approx\,  7.4 \times 10^{-4}\ {\rm W\,m^{-1}K^{-1}},\\
  k(z\approx 1{\rm\ m})&\,\approx\,  3.4 \times 10^{-3}\ {\rm W\,m^{-1}K^{-1}},\\
  k(z=0.08{\rm\ m})&\,\approx\,  (1.15-1.24) \times 10^{-2}\ {\rm W\,m^{-1}K^{-1}}\ \ (69.373{}^\circ~{\rm S},\,32.319{}^\circ~{\rm E}).
\end{align}
  For a given region (Table~\ref{tab:regional}), choose $\{\rho_0,\rho_\infty,z_c\}$, then solve for $(k_0,\ell,A)$ that best fit these anchors
  at the local temperature; carry the resulting posterior intervals into design margins.
\end{itemize}

\subsection{Temperature--illumination scalings and regional contrasts}
\label{sec:T-illumination}
The diurnal temperature wave couples into both thermophysical and electrical regimes. 
Define the thermal diffusivity $\kappa=k/(\rho c)$ and the frequency $\omega=2\pi/P$ with $P=2.55\times10^{6}$~s (synodic day).
The diurnal skin depth is
\begin{equation}
\delta \;\equiv\; \sqrt{\frac{2\kappa}{\omega}}
            \;=\; \sqrt{\frac{2\,k}{\rho\,c\,\omega}},
\end{equation}
which is typically $\delta\simeq 2$--$5$~cm for $k\simeq (0.7\text{--}3.4)\!\times\!10^{-3}$~W\,m$^{-1}$\,K$^{-1}$,
$\rho\simeq (1.3\text{--}1.6)\times10^{3}$~kg\,m$^{-3}$, and $c_p\simeq 600$--800~J\,kg$^{-1}$\,K$^{-1}$, matching Diviner-derived ranges.\footnote{See Sec.~\ref{sec:cond-thinert} and Fig.~\ref{fig:kdepth} for $k(z)$ and its $T$ dependence via the radiative term.} Within $z\lesssim\delta$, $k(T,\rho)=k_0(\rho/\rho_s)^\ell + A T^3$ (from Eq.~(\ref{eq:k-twochannel})) produces a day--night asymmetry in $\Gamma=\sqrt{k\rho c_p}$ 
and sets the temperature-phase lag.

The electrical repercussions follow from the conductivity of the pore/grain network, $\sigma(T)$, entering the charge-relaxation time $\tau_{\rm relax} \simeq \varepsilon_0 \varepsilon'/\sigma$ (Sec.~\ref{sec:dielec-losst}). For representative values $\varepsilon'\simeq 3$ and $\sigma \sim \{10^{-13}$, $10^{-15}$, $10^{-17}$\}~S\,m$^{-1}$, $\tau_{\rm relax}\simeq \{2.7\times 10^{2}$, $2.7\times 10^{4}$, $2.7\times 10^{6}$\}~s, respectively; thus colder, more porous environments (night, high latitudes, PSRs) maintain charge patches and field gradients far longer than sunlit equatorial terrains.

The dielectric response exhibits modest densification- and composition-driven changes (Looyenga mixing, Eq.~(\ref{eq:Looyenga})), 
with $\varepsilon'$ increasing $\sim 10$--30\% as $\rho$ rises from $1.4$ to $1.9$~g\,cm$^{-3}$ at fixed composition (CE-4 LPR trends), see Sec.~\ref{sec:dielec-losst}.
Loss tangent $\tan\delta$ increases with ilmenite fraction and temperature over the near-surface range relevant to LPR bands, 
implying larger RF attenuation in warm, Fe/Ti-rich maria than in cooler, feldspathic highlands. 

\paragraph{Regional contrasts (design priors):}
(1) {Equator, sunlit:} higher $T$ $\Rightarrow$ larger $A T^3$ in $k(T,\rho)$, smaller $\tau_{\rm relax}$, weaker/persistently smaller $|E|$ in uniform illumination.
(2) {Terminator/shadow edges:} sharp $E(z)$ gradients over decimeters; fields $10^2$--$10^3$~V\,m$^{-1}$ (Sec.~\ref{sec:charg-sheath-lift}).
(3) {PSRs and high latitudes:} $T\simeq 25$--40~K on PSR floors $\Rightarrow$ long $\tau_{\rm relax}$, favoring persistent charge segregation and large $|E|$ at boundaries.
(4) {Near- vs far-side:} composition differs statistically (far-side more feldspathic on average), biasing $\varepsilon'$ lower and $\tan\delta$ smaller;  this should be combined with local rock abundance and packing state when interpreting LPR and thermal data (Tables~\ref{tab:mechanism-observables}--\ref{tab:regional}).

\subsection{Vertical structure of the upper 3 meters (by region)}
\label{sec:upper3m}

We extend the density profile in Eq.~(\ref{eq:rhoz}) to $z\le 3$ m and make the implied porosity, dielectric constant, thermal conductivity, and overburden stress explicit. This provides design-ready priors for maria, highlands, pyroclastics, swirls, and PSR environments consistent with Apollo cores, Diviner inversions, and CE-4 LPR 
in the \emph{upper meters} \cite{Carrier1991,Hayne2017,Dong2021,Giannakis2021}. 

\paragraph{Density and porosity.}
With Eq.~(\ref{eq:rhoz}), porosity follows from $n(z)=1-\rho(z)/\rho_{s}$ using $\rho_{s}$ from composition (typically $\rho_{s}\sim 2.9$--$3.2$ g cm$^{-3}$). For the canonical parameters 
$\rho_{0}=1.4$ g cm$^{-3}$, $\rho_{\infty}=1.85$ g cm$^{-3}$, and $z_{c}=0.15$--$0.30$ m, $n(z)$ decreases from $\sim 0.5$--$0.55$ at the surface to $\sim 0.38$--$0.42$ by $z\simeq 1$ m and asymptotes slowly below $z\simeq 2$ m (consistent with Apollo \& Diviner). See Eq.~(\ref{eq:rhoz}) and Secs.~\ref{sec:den-poro} and \ref{sec:cond-thinert}. 

\paragraph{Vertical effective stress.}
The overburden to depth $z$ is (see Eq.~(\ref{eq:eff-stress}))
\begin{equation}
\sigma_{v}(z)=g_{\rm Moon}\int_{0}^{z}\rho(z')\,dz'
= g_{\rm Moon}\Bigl[\rho_{\infty}z-(\rho_{\infty}-\rho_{0})\,z_{c}\bigl(1-e^{-z/z_{c}}\bigr)\Bigr],
\label{eq:sigma_v}
\end{equation}
which gives $\sigma_{v}\!\approx\!2.8$ kPa at $1$ m and $\approx\!8.8$ kPa at $3$ m for the canonical parameters. This stress scale is consistent with the observed increase of $c$ with depth (Sec.~\ref{sec:shear}). 

\paragraph{Dielectric profile to 3 m.} Using Looyenga mixing Eq.~(\ref{eq:Looyenga}) with voids treated as $\varepsilon=1$ and solid volume fraction $1-n(z)$,
\begin{equation}
\varepsilon'^{1/3}(z)=(1-n(z))\,\varepsilon_{s}^{1/3}+n(z)\cdot 1,
\label{eq:eps_z}
\end{equation}
where $\varepsilon_{s}$ depends on composition (ilmenite-rich basalts vs. anorthosites). Eq.~(\ref{eq:eps_z}) combined with (\ref{eq:sigma_v}) reproduces the CE-4 LPR trend of increasing $\varepsilon'$ with depth/densification in the \emph{upper meters} and supports the regional ranges in Sec.~\ref{sec:dielec-losst}, see \cite{Dong2021,Giannakis2021}.

\paragraph{Thermal conductivity to 3 m.}
With Eq.~(\ref{eq:k-twochannel}),
\begin{equation}
k(z,T)=k_{0}\!\Big(\frac{\rho(z)}{\rho_{s}}\Big)^{\ell}+A\,T^{3},
\label{eq:k_z}
\end{equation}
whose $z$-dependence inherits Eq.~(\ref{eq:rhoz}). Calibrate $\{k_{0},\ell,A\}$ by matching Diviner night/day $k$ in the upper 
decimeters and the ChaSTE point at $z=0.08$ m at 69.373$^\circ$\,S (Sec.~\ref{sec:cond-thinert}). This yields a self-consistent $k(z)$ 
through $3$ m for thermal design. 

\paragraph{Regional priors (0--3 m).}
Adopt region-specific $(\rho_{0},\rho_{\infty},z_{c})$ consistent with Table~\ref{tab:regional}: 
maria $(1.40,1.85,0.15$--$0.22)$, highlands $(1.30,1.80,0.20$--$0.30)$, pyroclastics $(1.40,1.85,0.15$--$0.22)$, 
swirls $(1.40,1.85,0.15$--$0.25)$, PSR floors $(1.35,1.80,0.20$--$0.30)$. 
Combine Eq.~(\ref{eq:eps_z}) with composition-specific $\varepsilon_{s}$ (ilmenite fraction) to propagate into 
$\varepsilon'(z)$; then use Eq.~(\ref{eq:k_z}) for $k(z,T)$, anchoring to Diviner and ChaSTE. These profiles are directly compatible with the LPR $\varepsilon'$ ranges and the ``upper meters'' thermophysics 
summarized in Secs.~\ref{sec:thermphys-prop}--\ref{sec:elec-dielec-prop}. 

\section{Electrical and dielectric properties}
\label{sec:elec-dielec-prop}

The electrical and dielectric state follows from composition, porosity, illumination, and plasma forcing. We gather permittivity ranges, sheath fields, and charging pathways and update them with recent dayside H$^{-}$ detections.

\subsection{Dielectric constant and loss tangent}
\label{sec:dielec-losst}

Shallow regolith exhibits $\varepsilon' \approx 2.5$--3.5 with $\tan\delta \approx 10^{-3}$--$10^{-2}$; both rise with density and ilmenite fraction \citep{Dong2021,Giannakis2021}. For compositional mixing we use the Looyenga relation for the effective dielectric constant  \citep{Looyenga1965},
\begin{equation}
\varepsilon_{\rm eff}^{1/3}=\sum_{i} v_{i}\,\varepsilon_{i}^{1/3},
\label{eq:Looyenga}
\end{equation}
where $v_{i}$ and $\varepsilon_{i}$ are the volume fraction and dielectric constant of component $i$. At fixed composition, densification from $1.4$ to $1.9$ g cm$^{-3}$ raises $\varepsilon'$ by  $\approx 10$--$30$ \%, consistent with LPR trends \cite{Dong2021,Giannakis2021}.
Unless otherwise specified, reported $\varepsilon'$ and $\tan\delta$ refer to the upper meters at decimeter--meter wavelengths (LPR bands, cf. Sec.~\ref{sec:upper3m}) and near-surface temperature ranges.

\paragraph{Temperature and frequency dependence.}
Both $\varepsilon'$ and $\tan\delta$ vary with temperature, frequency, porosity, and composition. Laboratory UHF--SHF measurements of representative lunar end-members at fixed porosity show that cooling from $20\,^{\circ}$C to $-60\,^{\circ}$C (293$\rightarrow$213~K) reduces $\varepsilon'$ by $\sim$6--18\% (composition dependent) \cite{Kobayashi2023TempPerm}. Over 213--293~K, an engineering-linearized sensitivity $\alpha_{\varepsilon}\equiv -\tfrac{1}{\varepsilon'}\tfrac{\partial\varepsilon'}{\partial T}\sim (0.7$--$2.3)\times10^{-3}\,\mathrm{K^{-1}}$ captures this range to first order; extrapolation below $\sim$200~K (PSRs) remains poorly constrained and should be treated as an additional systematic uncertainty.

\paragraph{Charge relaxation time.}
The local charge-relaxation timescale is $\tau_{\mathrm{relax}}\simeq \varepsilon_{0}\varepsilon'/\sigma$, 
where $\sigma$ is the effective conductivity of the pore/grain network. 
Low $T$ and high porosity in polar and PSR environments imply small $\sigma$ and hence large $\tau_{\mathrm{relax}}$, 
biasing the sheath toward persistent patches and large field gradients along illumination boundaries. (For example, with $\varepsilon'=3$ and $\sigma=10^{-13},10^{-15},10^{-17}$~S\,m$^{-1}$, $\tau_{\mathrm{relax}}\approx 2.7\times 10^{2},\,2.7\times 10^{4},\,2.7\times 10^{6}$~s, respectively, see Table~\ref{tab:sigma_tau}) This provides the link from $k(T,\rho)$ Eq.~(\ref{eq:k-twochannel}) to the electrical regime (this section).

\paragraph{Neutral exosphere.} 
The neutral exosphere (He, Ne, Ar, Na/K, H$_2$/H) is collisionless at the surface with scale heights $\sim\!10^2$--$10^3$~km 
and near-surface densities $\ll 10^{10}\,{\rm m^{-3}}$; it is dynamically decoupled from grain motion in the 0--3~m layer, 
but sources (sputtering, micrometeoroid impact vapor, desorption) co-vary with processes that set dust charging and npFe$^0$ production. 
We therefore treat the exosphere as boundary forcing for charging (Sec.~\ref{sec:elec-dielec-prop}B) and dust maintenance (Sec.~\ref{sec:mobilization-transport}A).

\begin{table}[h]
\centering
\caption{Charge-relaxation priors by thermal/porosity regime (illustrative). Values assume $\varepsilon' \approx 3$ so that $\tau_{\rm relax}\!\approx\!\varepsilon_0\varepsilon'/\sigma$.}
\label{tab:sigma_tau}
\begin{tabular}{lcc}
\hline
Regime & $\sigma$ (S m$^{-1}$) & $\tau_{\rm relax}$ (s) \\
\hline\hline
Warm, Fe/Ti-rich maria (sunlit) & $10^{-13}$ & $2.7\times 10^{2}$ \\
Temperate highlands & $10^{-15}$ & $2.7\times 10^{4}$ \\
Cold, porous PSR / high latitudes & $10^{-17}$ & $2.7\times 10^{6}$ \\
\hline
\end{tabular}
\end{table}

\paragraph{Design note:} Use Table~\ref{tab:sigma_tau} with the $k(T,\rho)$ state from Sec.~\ref{sec:thermphys-prop} to pick a $\sigma$ consistent with $T$ and porosity; propagate $\tau_{\rm relax}$ into the time window over which $q$ can be treated quasi-constant in Eqs.~(\ref{eq:h_ball})--(\ref{eq:hover}).

\subsection{Charging, sheath fields, and lift including cohesion}
\label{sec:charg-sheath-lift}

Photoemission current density $J_{\rm ph}\approx e Y F_{\gamma}$ balances plasma currents to set floating potentials. Near-surface electric fields $E$ at terminators and in shadows reach $10^{2}$--$10^{3}$ V m$^{-1}$ \citep{Colwell2007,Stubbs2006}. Neglecting cohesion, the largest grain radius that can be lifted by an electric field $E$ at grain potential $\phi_{g}$ is
\begin{equation}
a_{\max}(E) = \left(\frac{3\,\varepsilon_{0}\,\phi_{g}\,E}{\rho_{p}\,g_{\mathrm{Moon}}}\right)^{1/2}.
\label{eq:a-noadh}
\end{equation}
For $\phi_{g}=5\ \mathrm{V}$, $E=300\ \mathrm{V\,m^{-1}}$, $\rho_{p}=3100\ \mathrm{kg\,m^{-3}}$, $g_{\mathrm{Moon}}=1.62\ \mathrm{m\,s^{-2}}$, Eq.~(\ref{eq:a-noadh}) gives $a_{\max}\approx 2.8\ \mu\mathrm{m}$.

\begin{figure}[t]
\centering
\begin{tikzpicture}
\begin{axis}[
  width=0.82\linewidth, height=6.0cm,
  xlabel={Grain radius $a$ ($\mu$m)},
  ylabel={Required field $E$ (V m$^{-1}$)},
  xmode=log, ymode=log,
  xmin=0.05, xmax=10,
  ymin=0.1,  ymax=1e12,
  grid=both,
  tick label style={/pgf/number format/fixed},
  legend pos=north east,
  legend cell align=left
]

% ---- No-adhesion component (circles; same nodes reused for E_adh markers)
\addplot+[thick, mark=o]
  coordinates {
    (0.05, 9.46e-2)   (0.10, 3.784e-1) (0.20, 1.5136)
    (0.50, 9.46)      (1.00, 3.784e1)  (2.00, 1.5136e2)
    (5.00, 9.46e2)    (10.0, 3.784e3)
  };
\addlegendentry{$E_{\mathrm{noadh}}(a)$}

% ---- Adhesion component (dashed line)
\addplot+[thick, dashed, mark=triangle*]
  coordinates {
    (0.05, 8.472e10) (0.10, 4.236e10) (0.20, 2.118e10)
    (0.50, 8.472e9)  (1.00, 4.236e9)  (2.00, 2.118e9)
    (5.00, 8.472e8)  (10.0, 4.236e8)
  };
\addlegendentry{$E_{\mathrm{adh}}(a)$}

% ---- Adhesion markers: small red triangles at the SAME x-points as the circles above
\addplot+[only marks, mark=triangle*, mark size=2.2pt,
          mark options={solid,fill=red,draw=red}, forget plot]
  coordinates {
    (0.05, 8.472e10) (0.10, 4.236e10) (0.20, 2.118e10)
    (0.50, 8.472e9)  (1.00, 4.236e9)  (2.00, 2.118e9)
    (5.00, 8.472e8)  (10.0, 4.236e8)
  };

% ---- Sum (dominantly adhesion in this range)
\addplot+[thick, mark=none]
  coordinates {
    (0.05, 8.472e10) (0.10, 4.236e10) (0.20, 2.118e10)
    (0.50, 8.472e9)  (1.00, 4.236e9)  (2.00, 2.118e9)
    (5.00, 8.472e8)  (10.0, 4.236e8)
  };
\addlegendentry{$E_{\mathrm{req}}(a)$}

\end{axis}
\end{tikzpicture}

\caption{Electric field required to lift a grain of radius $a$ when the charge is limited by $q\approx 4\pi\varepsilon_{0}a\phi_{g}$. Curves show the no-adhesion term $E_{\mathrm{noadh}}(a)$, the adhesion term $E_{\mathrm{adh}}(a)$ for $W=0.1$~J\,m$^{-2}$ and $R=5~\mu$m using the JKR pull-off
$F_{\mathrm{adh}}=\tfrac{3}{2}\pi W R$, and their sum $E_{\mathrm{req}}(a)$. Small red triangles mark
$E_{\mathrm{adh}}(a)$ at the same abscissae where open circles mark $E_{\mathrm{noadh}}(a)$. Across the micron range, adhesion
dominates and $E_{\mathrm{req}}$ far exceeds typical sheath values ($10^{2}$--$10^{3}$~V\,m$^{-1}$) unless grains are pre-loosened. Also, adhesion is made explicit via $C_{\rm adh}$, turning electrostatic lift into a pre-liberation criterion rather than a field--only condition.}
\label{fig:E_vs_a_required}
\end{figure}
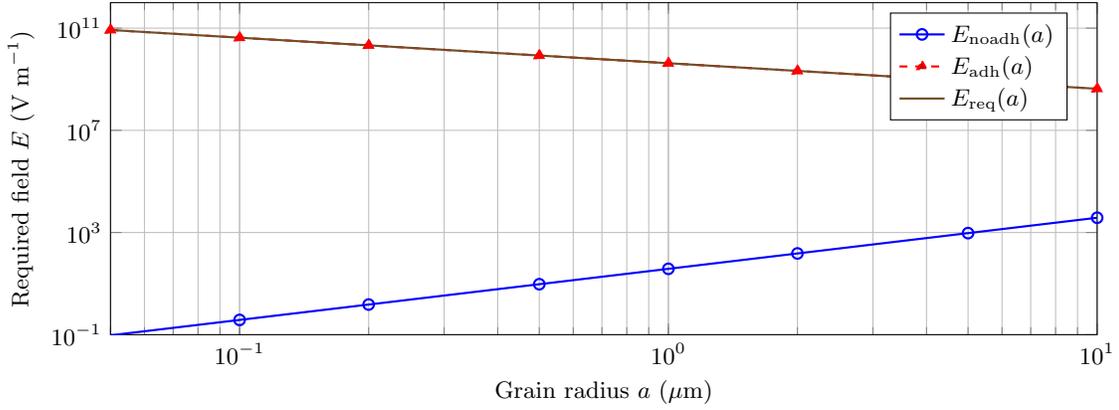

Including adhesion (JKR form at a single asperity of radius $R$ with work of adhesion $W$, see Eq.~(\ref{eq:JKR-pull-off})),
\begin{equation}
qE \gtrsim m g_{\mathrm{Moon}} + F_{\mathrm{adh}}, \qquad
F_{\mathrm{adh}} \approx \tfrac{3}{2}\pi W R,\qquad
q=4\pi\varepsilon_{0}a\phi_{g},
\label{eq:a-adh}
\end{equation}
so that the field required to separate a grain of radius $a$ is
\begin{equation}
E_{\mathrm{req}}(a) \gtrsim \frac{\tfrac{4}{3}\pi a^{3}\rho_{p}g_{\mathrm{Moon}} + \tfrac{3}{2}\pi W R}{4\pi\varepsilon_{0}a\phi_{g}}=\Big(\frac{\rho_{p}g_{\mathrm{Moon}}}{3\varepsilon_{0}\phi_{g}}\Big)a^2+\Big(\frac{3 W R}{8\varepsilon_{0}\phi_{g}}\Big)\frac{1}{a}
\qquad \Rightarrow \qquad E_{\mathrm{req}}(a) \gtrsim C_{\rm no}\,a^2+C_{\rm adh}\,\frac{1}{a}.
\label{eq:E-adh}
\end{equation}
With $W=0.1~\mathrm{J\,m^{-2}}$ and $R=5~\mu\mathrm{m}$, 
$F_{\mathrm{adh}}=\tfrac{3}{2}\pi W R \simeq 2.36~\mu\mathrm{N}$, which exceeds the weight of a $1~\mu\mathrm{m}$ grain by $\sim 1.1\times 10^{8}$ (see Fig.~\ref{fig:E_vs_a_required}). This explains why pre-liberation (vibration, micro-impacts, or gas shear) is commonly required prior to sheath acceleration.

Even if rare transients were to push near-surface fields toward $10^{4}$--$10^{5}$~V\,m$^{-1}$ in special geometries, (\ref{eq:E-adh}) still implies $E_{\mathrm{req}}\!\gg\!10^{7}$~V\,m$^{-1}$ for micron-scale grains under the reference $(W,R)$, so pre-liberation remains necessary even then.

For representative values $W=0.1~\mathrm{J\,m^{-2}}$, $R=5\times 10^{-6}~\mathrm{m}$, $\rho_{p}=3.1\times 10^{3}~\mathrm{kg\,m^{-3}}$, $g_{\mathrm{Moon}}=1.62~\mathrm{m\,s^{-2}}$, $\varepsilon_{0}=8.854\times 10^{-12}~\mathrm{F\,m^{-1}}$, and $\phi_{g}=5~\mathrm{V}$, from (\ref{eq:E-adh}), we have\footnote{$C_{\rm no}$ has units of [\si{\per\meter\squared}] and $C_{\rm adh}$ of [\si{\volt}], so that $C_{\rm adh}/a$ has units of \si{\volt\per\meter}. With $a$ in \si{\meter}, Eq.~(\ref{eq:E-adh}) yields $E$ in \si{\volt\per\meter}.}
\begin{equation}
C_{\rm no}=\frac{\rho_{p}g_{\mathrm{Moon}}}{3\varepsilon_{0}\phi_{g}} = 3.781\times 10^{13}~ {\rm m}^{-2} \qquad {\rm and} \qquad 
C_{\rm adh}=\frac{3 W R}{8\varepsilon_{0}\phi_{g}} = 4.236\times 10^{3}~~ ({\textrm{SI, $a$ in m}}).
\label{eq:C_no-adh}
\end{equation}
As a result, Eq.~(\ref{eq:E-adh}) yields fields far exceeding realistic sheath values unless grains are pre-loosened by vibration, micro-impacts, thermal cycling, or gas shear. For example, $a=1~\mathrm{\mu m}$ gives $E_{\mathrm{req}} \sim 4.2\times 10^{9}~\mathrm{V\,m^{-1}}$, dominated by $F_{\mathrm{adh}}$; at $a=5~\mathrm{\mu m}$ the requirement is still $E_{\mathrm{req}} \sim 1.0\times 10^{9}~\mathrm{V\,m^{-1}}$. This quantifies why electrostatic mobilization of adhered grains requires pre-liberation even when $E \sim 10^{2}$--$10^{3}~\mathrm{V\,m^{-1}}$ near terminators and shadow edges (see Table~\ref{tab:adhesion} for numerical examples and Fig.~\ref{fig:E_vs_a_required} for the plotted behavior.) 

\begin{table}[h]
\centering
\caption{Illustrative adhesion-aware lift thresholds from (\ref{eq:E-adh}) with $\rho_p=\SI{3100}{kg.m^{-3}}$, $g_{\rm Moon}=\SI{1.62}{m.s^{-2}}$, $\phi_g=\SI{5}{V}$,
$W=\SI{0.1}{J.m^{-2}}$, $R=5~\mu$m, and $F_{\rm adh}=\tfrac{3}{2}\pi W R$.
Values highlight the dominance of adhesion over gravity for micron-scale grains.}
\label{tab:adhesion}
\small
\begin{tabular}{lccc}
\hline
Radius $a$ & $m g_{\mathrm{Moon}}$ (N) & $F_{\mathrm{adh}}$ (N) & $E_{\mathrm{req}}$ (V m$^{-1}$) \\
\hline\hline
$0.1~\mathrm{\mu m}$ & $2.10\times 10^{-17}$ & $2.36\times 10^{-6}$ & $\sim 4.24\times 10^{10}$ \\
$1.0~\mathrm{\mu m}$ & $2.10\times 10^{-14}$ & $2.36\times 10^{-6}$ & $\sim 4.24\times 10^{9}$  \\
$5.0~\mathrm{\mu m}$ & $2.63\times 10^{-12}$ & $2.36\times 10^{-6}$ & $\sim 8.47\times 10^{8}$  \\
\hline
\end{tabular}
\end{table}

\paragraph{Design-relevant optimum of the adhesion--gravity trade.}
Because $E_{\rm req}(a)=C_{\rm no}a^2+C_{\rm adh}/a$ [Eq.~\eqref{eq:E-adh}], the minimum field occurs at
\begin{equation}
a_\ast=\left(\frac{C_{\rm adh}}{2C_{\rm no}}\right)^{1/3},
\qquad
E_{\rm req,min}=E_{\rm req}(a_\ast)=\frac{3}{2^{2/3}}\,C_{\rm no}^{1/3}C_{\rm adh}^{2/3}.
\label{eq:Ereq_min}
\end{equation}
For the reference values used in Fig.~\ref{fig:E_vs_a_required} ($\phi_g=5~\mathrm{V}$, $\rho_p=3100~\mathrm{kg\,m^{-3}}$,
$W=0.1~\mathrm{J\,m^{-2}}$, $R=5~\mu\mathrm{m}$), this gives
$a_\ast\simeq 3.8\times 10^{-4}~\mathrm{m}$ and $E_{\rm req,min}\simeq 1.7\times 10^{7}~\mathrm{V\,m^{-1}}$,
i.e., even the most ``lift-favorable'' size under these adhesion assumptions still requires fields far above the
$10^{2}$--$10^{3}~\mathrm{V\,m^{-1}}$ envelope in Table~\ref{tab:adhesion}; therefore bed-mediated detachment and/or mechanical pre-liberation
remain necessary for micron-scale grains.

\paragraph{Bed-mediated detachment: microcavity charge pumping and the patched-charge model (PCM).}
Eqs.~(\ref{eq:a-adh})--(\ref{eq:E-adh}) treat a grain as an isolated sphere and therefore provide a \emph{lower bound} on the external field needed to lift an already-detached (or very weakly adhered) particle. In contrast, a series of laboratory experiments has demonstrated that dust resting in a porous, insulating \emph{bed} can be self-ejected under UV and/or plasma exposure due to charge pumping in microcavities between grains \citep{Wang2016DustChargingTransport,Schwan2017ChargeState,Hood2018RateLofting}. In the PCM picture, photoelectrons (and/or secondary electrons) emitted into microcavities are preferentially re-absorbed by surrounding grains, producing large \emph{negative} grain charges and strong inter-particle repulsion even under overall dayside illumination \citep{Wang2016DustChargingTransport}.

A simple order-of-magnitude detachment criterion for two like-charged grains is obtained by comparing Coulomb repulsion to the adhesive pull-off force. Approximating the repulsive interaction at a characteristic separation $d$,
\begin{equation}
F_{\rm rep}\sim \frac{q_{\rm p}^{2}}{4\pi\varepsilon_{0}d^{2}}\;\gtrsim\;F_{\rm adh},
\label{eq:repulsion_vs_adhesion}
\end{equation}
yields a required patch/particle charge
\begin{equation}
q_{\rm p,req}\;\gtrsim\;d\,\sqrt{4\pi\varepsilon_{0}F_{\rm adh}}
\;=\;\pi d\,\sqrt{6\varepsilon_{0} W R},
\label{eq:q_patch_req}
\end{equation}
where $F_{\rm adh}$ is given by Eq.~(\ref{eq:JKR-pull-off}). For the reference values $W=\SI{0.1}{J.m^{-2}}$ and $R=5~\mu$m (so $F_{\rm adh}\simeq 2.36~\mu$N), Eq.~(\ref{eq:q_patch_req}) gives
\begin{equation}
\frac{q_{\rm p,req}}{e}\;\approx\;1.0\times 10^{3}\left(\frac{d}{10~{\rm nm}}\right)
\left[\left(\frac{W}{0.1~{\rm J\,m^{-2}}}\right)\left(\frac{R}{5~\mu{\rm m}}\right)\right]^{1/2},
\label{eq:q_patch_req_numeric}
\end{equation}
showing that nanometer-scale microcavity separations can, in principle, enable electrostatic detachment at charge levels far below what would be required if the interaction acted only over grain-scale distances.

Consistent with this mechanism, lofted grains have been reported to carry unusually large \emph{negative} charges (up to $10^{5}$--$10^{6}$ electrons for $\sim 20~\mu$m-radius grains) and to exhibit peak lofting rates of order $\sim 5~{\rm cm^{-2}\,s^{-1}}$ in representative 1~AU conditions \citep{Schwan2017ChargeState,Hood2018RateLofting}. Therefore, Eq.~(\ref{eq:E-adh}) should be interpreted as the external-field requirement for lifting an \emph{isolated adhered grain}; bed-scale microcavity charging can provide an additional detachment pathway that bypasses the need for purely mechanical pre-liberation in some regimes.

\subsection{Triboelectric charging}

Contact/frictional charging among insulating fines and with rover or suit materials is efficient in vacuum; net charge depends on work-function offset, contact area, and history \citep{Forward2009,Sternovsky2002}. Tribocharge increases adhesion and modifies separation thresholds.

\paragraph{Vehicle charging and ESD posture (design note).}
Given $\tau_{\rm relax} \simeq \varepsilon_{0}\varepsilon'/\sigma$ (Sec.~\ref{sec:elec-dielec-prop}), cold, porous PSR and high--latitude soils exhibit long relaxation times, admitting quasi--static patch potentials on vehicle and local soil. 
Near shadow/illumination boundaries, fields $E_0\sim 10^{2}$--$10^{3}$~V\,m$^{-1}$ (Sec.~\ref{sec:charg-sheath-lift}) can persist over $\lambda\sim$\,few--tens cm, consistent with the adhesion--aware lift thresholds (Fig.~\ref{fig:E_vs_a_required}). 
Design implications: (i) reference chassis/booms to a single ground and avoid floating appendages in strong gradients;
(ii) pair electrostatic dust mitigation with \emph{mechanical pre--liberation} where relevant, \emph{and} account for bed-mediated electrostatic ejection via microcavity charge pumping (patched-charge model) as an additional detachment pathway under UV/plasma exposure \citep{Wang2016DustChargingTransport,Schwan2017ChargeState,Hood2018RateLofting}; and (iii) treat dayside H$^{-}$ (Sec.~\ref{sec:elec-dielec-prop}) as a modifier to sheath composition, particularly for daylight traverses over anomaly boundaries.

\subsection{Near-surface negative ion layer (dayside)}
\label{sec:near-surf-ion}

Chang'e-6 NILS detected a dayside near-surface H$^{-}$ population over the far-side, with $(2.5^{+1.2}_{-0.8})$ percent of incident solar-wind protons charge-exchanged and backscattered as H$^{-}$; a local number density $\sim 1.8\times 10^{-1}~\mathrm{cm^{-3}}$ and a scale height of order $10~\mathrm{km}$ were inferred, limited by rapid photo-detachment \cite{Wieser2025NILSHminus}.  The presence of negative ions modifies photoelectron sheath structure, grain-charging time constants, and therefore the spatio-temporal envelope in which Eq.~(\ref{eq:E-adh}) can be satisfied. Practical modeling should include electrons, positive ions, and H$^{-}$ with photo-detachment to assess daylight lofting near boundaries and over magnetic anomalies.

\paragraph{Link to optical maturity.} The same processes that drive charge balance (photoelectron production, implantation, micrometeoroid impact vaporization) also control the rate at which npFe$^{0}$ accumulates. Consequently, optical maturity metrics and CF shifts provide an independent clock on the surface exposure that, in principle, can be fused with charging models to predict where and when electrostatic mobilization is favored. Sec.~\ref{sec:opt-prop} summarizes the relevant spectral diagnostics.

\paragraph{Neutral exosphere context.}
The neutral lunar exosphere is collisionless on meter scales; neutral number densities and mean free paths imply that drag over the $\lesssim$\,2--3\,s timescales of sub-meter hops is negligible compared to electrostatic and gravitational forces considered here. Consequently, the dynamical budgets in Eqs.~(\ref{eq:a-noadh})--(\ref{eq:E-adh}) and (\ref{eq:hover}) are governed by the charged component (photoelectrons, ions, and the dayside near-surface H$^{-}$ detected by NILS) rather than by neutral drag.

\paragraph{Measured properties and loss-limited scale height.}
NILS reports H$^{-}$ energies in the $\sim 50$--$400$~eV range and interprets the vertical distribution as loss-limited by photo-detachment, with a characteristic lifetime $\tau_{\rm pd}\sim 70$~ms and a scale height of order $H_{H^-}\sim 10$~km \citep{Wieser2025NILSHminus}. A simple consistency check follows from ballistic transport with photo-detachment loss:
\begin{equation}
H_{H^-}\;\sim\;\langle v_{z}\rangle \tau_{\rm pd}\;\approx\;\frac{v_{H^-}}{2}\tau_{\rm pd},
\label{eq:Hminus_scaleheight}
\end{equation}
where $v_{H^-}\!=\!\sqrt{2E_{H^-}/m_{H}}$. For $E_{H^-}\sim 239$~eV and $\tau_{\rm pd}\sim 70$~ms, $v_{H^-}\tau_{\rm pd}\sim 15$~km, consistent with the inferred $\sim$10~km scale height \citep{Wieser2025NILSHminus}.

\paragraph{Electronegativity and its impact on the sheath length scale.}
A convenient non-dimensional measure of the negative-ion importance is the electronegativity
\begin{equation}
\alpha(z)\equiv \frac{n_{H^-}(z)}{n_e(z)} .
\label{eq:alpha_electronegativity}
\end{equation}
In a linearized (small-amplitude) multi-species limit, the shielding scale entering exponential closures like in Eq.~\eqref{eq:E-integral} is controlled by the generalized Debye length,
\begin{equation}
\lambda_D^{-2} \;\approx\; \frac{e^2}{\epsilon_0 k_B}\left(\frac{n_e}{T_e}+\frac{n_i}{T_i}+\frac{n_{H^-}}{T_{H^-}}\right),
\label{eq:Debye_multi}
\end{equation}
so that, all else equal, increasing $n_{H^-}$ (or decreasing $T_{H^-}$) decreases $\lambda_D$ and steepens near-surface potential gradients. In the near-surface photoelectron layer $n_e$ can be large, so $\alpha\ll 1$ may imply only a small perturbation to $\lambda$ very close to the surface; however, because $n_e(z)$ typically falls rapidly with altitude while loss-limited $n_{H^-}(z)$ can persist over kilometer scales (Sec.~\ref{sec:near-surf-ion}.c--d), $\alpha(z)$ may increase with height, enabling negative space charge to reshape $\phi(z)$ outside the first meters (potential humps/wells), which can in turn modify dust trajectories that extend above the near-surface region (e.g., \cite{Franklin2002ElectronegativeDifferent,Riemann1991BohmCriterion,Whipple1981PotentialsSurfacesSpace}).

\paragraph{Minimal sheath model with an electronegative component.}
To first order, the near-surface potential must satisfy Poisson’s equation with electrons, positive ions, and negative ions,
\begin{equation}
\varepsilon_{0}\frac{d^{2}\phi}{dz^{2}}=-e\left(n_{i}(z)-n_{e}(z)-n_{H^-}(z)\right),
\label{eq:Poisson_e_i_Hminus}
\end{equation}
with $n_{H^-}(z)$ constrained by Eq.~(\ref{eq:Hminus_scaleheight}) in the loss-limited regime. Even if $n_{H^-}$ is small compared to the peak photoelectron density close to the surface, it can contribute negative space charge over kilometer scales and thereby alter the effective $\lambda$ and/or introduce non-monotonic potential structure (e.g., shallow potential wells/humps) relevant for dust trajectories that extend above the first meters.

\paragraph{Implications for dust charging and mobilization.}

Dust charging enters lift and transport scalings through the capacitance-limited charge
$q(a)=4\pi\varepsilon_{0}a\phi_{g}$ [Eq.~(\ref{eq:a-adh})] 
and the near-surface sheath model $E(z)=E_{0}\exp(-z/\lambda)$, with $\Delta V_{\rm sh}\simeq E_{0}\lambda$ given by Eq.~\eqref{eq:E-integral}. The floating potential $\phi_{g}$ is determined by current balance. With H$^{-}$ present, the net collected current includes an additional negative-ion term,
\begin{equation}
I_{\rm tot}(\phi_{g})=I_{e}(\phi_{g})+I_{i}(\phi_{g})+I_{\rm ph}(\phi_{g})+I_{H^-}(\phi_{g})+I_{\rm se}(\phi_{g})\;=\;0,
\label{eq:current_balance_with_Hminus}
\end{equation}
so that both the equilibrium $\phi_{g}$ and the charging time constant $\tau_{\rm ch}\sim C/(\partial I_{\rm tot}/\partial \phi_{g})$ (with $C\simeq 4\pi\varepsilon_{0}a$) are modified relative to the electron+ion+photoelectron case. Since Eq.~(\ref{eq:E-adh}) scales as $E_{\rm req}\propto 1/\phi_{g}$ for fixed $(W,R)$, any H$^{-}$-driven shift in $\phi_{g}$ maps directly into the detachment threshold, and any modification to $E(z)$ (through Eq.~(\ref{eq:Poisson_e_i_Hminus})) maps into $\Delta V_{\rm sh}$ and thus into $h_{\rm ball}$ and $h_{*}$ [Eqs.~(\ref{eq:h_ball})--(\ref{eq:hover})]. In practice, incorporating H$^{-}$ requires (i) enforcing Eq.~(\ref{eq:current_balance_with_Hminus}) for $\phi_{g}$ and $\phi_{\rm surface}$, (ii) updating the field priors $(E_{0},\lambda)$ in Eq.~(\ref{eq:E-integral}), and (iii) propagating the updated $q(a)$ into Eqs.~(\ref{eq:E-adh}), (\ref{eq:h_ball})--(\ref{eq:hover}).

\subsection{Surface-bounded neutral exosphere and coupling to dust}
\label{sec:exosphere-neutral}

The Moon hosts a surface-bounded, collisionless neutral exosphere sourced by photon-stimulated desorption (PSD),
solar-wind sputtering, micrometeoroid impact vaporization (MIV), and, for $^{40}$Ar, radiogenic outgassing
\citep{Stern1999RG,KillenIp1999}. Dominant species include He, Ne, Ar, H$_2$/H, and alkalis (Na, K)
\citep{Stern2012_GRL_LAMP_He,Benna2015He,Benna2019_NatGeosci_Water,Dhanya2021}. In a ballistic exosphere, the density scale height
\begin{equation}
H = \frac{k_{B} T_{s}}{m\,g_{\rm Moon}}
\label{eq:Hscale}
\end{equation}
is set by a characteristic source (kinetic) temperature $T_{s}$ and molecular mass $m$.
Representative values: $H_{\rm He}\sim 100$--$400$~km for $T_{s}=100$--$400$~K; $H_{\rm Ar}\sim 10$--$30$~km for
$T_{s}=100$--$300$~K; Na/K have $H\sim 100$--$200$~km owing to hot PSD/MIV source distributions
\citep{Stern1999RG,KillenIp1999,Benna2015He,Benna2019_NatGeosci_Water,Dhanya2021}. If a vertical column $N$ is known, a near-surface
density estimate follows as $n_{0}\simeq N/H$.

\paragraph{Variability and day--night/far--near differences.}
LADEE/NMS observed strong temporal and local-time variability, with He tracking solar-wind flux and
nightside enhancements, while $^{40}$Ar varies with longitude/local time consistent with a radiogenic source and
cold-trapping/release \citep{Benna2015He,Dhanya2021}. LRO/LAMP detections of He corroborate nightside
columns of order $10^{11}$--$10^{12}$~cm$^{-2}$ \citep{Stern2012_GRL_LAMP_He}.
Because the neutral mean free path is $\,\gg\,$~meters under all conditions, neutral drag is negligible for
0--3~m dust dynamics (Sec.~\ref{sec:near-surface-0-3m}); the exosphere influences dust primarily by
\emph{production} (PSD/sputtering/MIV that generate fines or alter coatings) and by providing independent tracers
of surface processes that also control maturity and charging.

\paragraph{Coupling to dust/charging.}
PSD/sputtering efficiencies scale with UV/EUV and ion fluxes that also set photoemission and sheath structure, so neutral columns (He, Na, K) can serve as environmental proxies for the likelihood of electrostatic mobilization in strongly forced geometries (terminators, shadow edges, PSR rims). In PSRs, low $T$ and long charge-relaxation times (Sec.~\ref{sec:elec-dielec-prop}) favor persistent fields; any local neutral release (e.g., MIV) is quickly ballistic and does not screen $E(z)$ in the first meters.

\paragraph{Design use.}
Use Eq.~(\ref{eq:Hscale}) with species mass and a site-specific $T_{s}$ to project $N\!\to\!n_{0}$ (or vice versa); interpret neutral variability in concert with $E(z)$ (Sec.~\ref{sec:charg-sheath-lift}) and the 0--3~m dust envelope (Sec.~\ref{sec:near-surface-0-3m}) to bound dust generation and transport.

\subsection{Context for dust-neutral exosphere design}
\label{sec:exosphere}

As we discussed, the Moon hosts a surface-bounded, collisionless neutral exosphere with species-dependent source and loss: noble gases (He, Ar, Ne) from solar wind implantation and radiogenic outgassing; alkalis (Na, K) from micrometeoroid impact vaporization and photon-stimulated desorption  \citep{Stern1999RG,Benna2015He}. Typical nightside number densities are $n_{\mathrm{He}}\!\sim\!10^{3}$--$10^{5}\,\mathrm{cm^{-3}}$ and
$n_{\mathrm{Ar}}(^{40}\mathrm{Ar})\!\sim\!10^{3}$--$10^{4}\,\mathrm{cm^{-3}}$, with scale heights $H = kT/(m g_{\mathrm{Moon}})$ ranging from tens of km (alkalis) to $\gtrsim 100$~km (He) for $T\sim 100$--$400$~K
\citep{Stern1999RG,Benna2015He}. Gas--grain drag is negligible for micron scales in vacuum, so the exosphere does not alter the kinematics of the 0--3~m dust hops derived here \eqref{eq:h_ball}--\eqref{eq:hover}. However, the exosphere modulates surface charging and photoemission context (through day/night $T$ and UV illumination), and it supplies Na/K that can form transient films on cold optical/thermal surfaces. For rover design, exosphere relevance enters via (i) day/night fouling rates (thin film deposition), and (ii) context for photoemission yields used in sheath models.

\section{Optical and maturity diagnostics}
\label{sec:opt-prop}

Lunar reflectance is non-Lambertian with a strong opposition surge. For particulate surfaces the Hapke bidirectional reflectance formalism \citep{Hapke2012} provides a compact, dust-relevant parameterization. For incidence $i$, emission $e$, and phase angle $g$ (with $\mu_{0}\!=\!\cos i$, $\mu\!=\!\cos e$), the radiance factor is
\begin{equation}
r(i,e,g) \;=\; \frac{w}{4}\,\frac{\mu_{0}}{\mu_{0}+\mu}\Big[\Big(1+B(g)\Big)\,P(g)\;+\;H(\mu_{0})H(\mu)-1\Big]\;S(\bar{\theta}),
\label{eq:hapke-r}
\end{equation}
where $w$ is the single-scattering albedo, $P(g)$ a particle phase function (often a 1- or 2-term Henyey--Greenstein), $B(g)=B_{\mathrm{SH}}(g)+B_{\mathrm{CB}}(g)$ the opposition term (shadow hiding + coherent backscatter), $H$ are the Chandrasekhar $H$-functions for multiple scattering, and $S(\bar{\theta})$ accounts for macroscopic roughness with mean slope $\bar{\theta}$. Eq.~\eqref{eq:hapke-r} may be used to link grain-scale properties (PSD, shape, npFe$^{0}$, SSA) to observables.

\subsection{Dust-controlled photometric behavior}
\label{sec:opt-photometry}

Angular, fine grains (high SSA) lower $w$ at visible wavelengths by increasing internal path length and absorption and enhance backscatter through larger $B_{0}$ (the $g\!\to\!0$ limit of $B(g)$). In the small-$w$ limit, the reflectance sensitivity to $w$ is approximately linear:
\begin{equation}
\left.\frac{\partial r}{\partial w}\right|_{w\ll 1}
\approx 
\frac{1}{4}\,\frac{\mu_{0}}{\mu_{0}+\mu}\Big[\big(1+B\big)P+H(\mu_{0})H(\mu)-1\Big]\,S(\bar{\theta}).
\label{eq:drdw}
\end{equation}
Thus, modest changes in $w$ from space weathering or dust coatings induce nearly proportional changes in $r$ once geometry is fixed. Mature maria have lower $w$ and stronger backscatter; highlands are brighter. Diviner CF plus NIR band-depths provide compositional and maturity constraints \citep{Lucey2017CF,Green2011}. Typical normal albedo ranges (0.55\,$\mu$m) are $\sim 6$--$10\%$ for maria and $\sim 12$--$18\%$ for highlands.  

\paragraph{Opposition terms as microtexture proxies.}
Shadow-hiding and coherent-backscatter parameters $(B_{0},\,h)$ covary with microtexture: higher fines/SSA $\Rightarrow$ larger $B_{0}$ and smaller $h$. Tracking $(B_{0},h)$ in repeat imaging (constant $i$, $e$) offers a non-contact indicator of near-surface textural change (e.g., plume dusting or sorting). Note that here $h$ denotes the angular half-width (in radians) of the opposition surge in Hapke's opposition functions in Eq.~(\ref{eq:hapke-r}):
$B_{\rm SH}(g)=B_{0,{\rm SH}}\bigl[1+(h_{\rm SH}^{-1}\tan(g/2))\bigr]^{-1}$ and 
$B_{\rm CB}(g)=B_{0,{\rm CB}}\bigl[1+(h_{\rm CB}^{-1}\tan(g/2))\bigr]^{-2}$.
Smaller $h$ implies a narrower, steeper surge near $g\!\to\!0^\circ$, diagnostic of finer, higher-SSA microtexture. Thus, $h$ is the shorthand for these width parameters ($h_{\rm SH}$ and/or $h_{\rm CB}$). Finer, more porous/rough microtexture (higher SSA) typically yields larger amplitude $B_0$ and smaller $h$ (a steeper, narrower opposition spike). On the Moon, indicative values are $h_{\rm SH}\sim0.02\text{--}0.08$~rad ($\approx 1\text{--}5^\circ$) and $h_{\rm CB}\sim 0.01\text{--}0.05$~rad, varying with maturity and texture.

\subsection{Space weathering, npFe$^{0}$, and maturity metrics}
\label{sec:opt-weathering}

Solar-wind implantation and micro-impacts generate npFe$^{0}$ in 10--200~nm rims and in agglutinitic glass, decreasing $w$, reddening continua, and suppressing mafic bands \citep{Lucey2006,Taylor2001}. A first-order attenuation of a band depth ${\tt BD}$ with solar-wind fluence $\phi_{\rm sw}$ was introduced in Sec.~\ref{sec:formation} and captures the maturation trend used for photometric/maturity correction.\footnote{Space-weathering rims (10--200 nm) and agglutinitic glass generate npFe$^{0}$,
lowering albedo, reddening continua, and suppressing mafic bands 
(see Eq.~(\ref{eq:BD}) in Sec.~\ref{sec:formation} for ${\tt BD}\propto e^{-\alpha \phi_{\rm sw}}$.)} 
The ferromagnetic resonance index $I_{\rm s}$/FeO and Diviner CF position covary with npFe$^{0}$ abundance and composition \citep{Morris1978,Lucey2017CF,Wang2017}, enabling cross-calibration of maturity between VNIR and TIR.

\subsection{Christiansen Feature (CF) in the thermal IR}
\label{sec:opt-CF}

The Diviner Christiansen Feature (CF; peak in emissivity where refractive index $\approx 1$) shifts with polymerization (felsic vs mafic), Fe/Ti content, and weathering state.  
To first order, the apparent CF position $\nu_{\rm CF}^{\rm app}$ inferred from radiance spectra depends on both composition and the thermal state via the Planck weighting:
\begin{equation}
L_{\nu}(\nu,T)\;=\;\epsilon_{\nu}(\nu)\,B_{\nu}(T),
\qquad 
\nu_{\rm CF}^{\rm app}\,=\,\arg\max_{\nu}\,\epsilon_{\nu}(\nu)B_{\nu}(T).
\label{eq:CF-app}
\end{equation}
Consequently, diurnal/seasonal $T$ variations (set by $k(T,\rho)$; Sec.~\ref{sec:thermphys-prop}) introduce small, geometry-dependent CF biases unless temperature is co-modeled. Combining CF with VNIR band depths separates basaltic (maria) from anorthositic (highlands) terrains and can be mapped against LPR-inferred $\varepsilon'$ to constrain composition and density jointly \citep{Green2011,Lucey2017CF}. 

\subsection{From dust to photometric correction (operational recipe)}
\label{sec:opt-correction}

For site-to-site or time-series comparisons (e.g., monitoring dust deposition near landers), normalize BRF to a standard geometry $(i_{0},e_{0},g_{0})$ by the Hapke ratio:
\begin{equation}
R_{\rm norm}(\lambda)\;=\;R_{\rm meas}(\lambda)\,
\frac{r(i_{0},e_{0},g_{0};\,\Theta)}{r(i,e,g;\,\Theta)},
\qquad
\Theta\equiv\{w(\lambda),P,B,\bar{\theta}\}.
\label{eq:norm}
\end{equation}
Using Eq.~\eqref{eq:norm} with a fixed $\Theta$ isolates geometry; allowing only $(B_0,h)$ (and optionally $w$) to vary (with $B(g)=B_{\rm SH}(g)+B_{\rm CB}(g)$ as in Eq.~\eqref{eq:hapke-r}) is an effective way to detect dust textural changes without conflating composition. Because npFe$^{0}$ modifies both $w(\lambda)$ and CF, joint VNIR+TIR normalization [Eq.~\eqref{eq:CF-app}+Eq.~\eqref{eq:norm}] mitigates maturity/composition cross-talk \citep{Lucey2017CF}. 

\subsection{Optical consequences for dust transport and system performance}
\label{sec:opt-impacts}

\emph{Diagnostics of near-surface dust:} Phase-angle dependence near $g\!\to\!0$ is acutely sensitive to surficial fines; a measurable increase in $(B_{0},\,r_{0})$ at fixed composition flags fresh coatings or fine sorting by electrostatic hopping (Sec.~\ref{sec:mobilization-transport}). LADEE's lack of a dense, persistent $0.1~\mu$m high-altitude layer is consistent with \emph{localized}, meter-scale hopping that does not build an optically thick haze aloft. 

\emph{Engineering impacts:} Dust coatings lower $w$, increase backscatter ($B_{0}\!\uparrow$), and reduce thermal IR contrast (via $\epsilon_{\nu}\!\to\!1$ near CF), degrading albedo-based power/thermal margins and instrument radiometry. These effects are detectable with co-registered VNIR/TIR observations using Eqs.~\eqref{eq:hapke-r}--\eqref{eq:CF-app}. Terrain differences (maria vs.\ highlands; near- vs.\ far-side) expressed in Fe/Ti and maturity (Tables~\ref{tab:regional}--\ref{tab:properties}) translate to systematic albedo, spectral slope, and CF contrasts, which should be treated as priors in contamination/discoloration assessments. 

To isolate dust-texture changes from composition and geometry,
we recommend (i) fixing $\bar\theta$ and using a reduced phase function to track  $(B_{0},h)$ at small phase angles, and (ii) normalizing TIR spectra about the CF, Eq.~(\ref{eq:CF-app}), while using co-registered VNIR band depths. This joint VNIR+TIR normalization mitigates maturity/composition cross-talk and is sensitive to surficial fine coatings expected from near-surface hopping (Sec.~\ref{sec:mobilization-transport}).

\paragraph{Dust coatings on engineered surfaces (power, thermal, RF).}
For a surface with solar absorptance $\alpha_{\rm s}$ and thermal emissivity $\varepsilon_{\rm s}$, partially covered by a dust film of area fraction $f_{\rm d}$ 
(with $\alpha_{\rm d}$, $\varepsilon_{\rm d}$ appropriate to local composition/maturity; Sec.~\ref{sec:opt-prop}), a first--order areal mixing gives
\begin{equation}
\alpha_{\rm eff} \approx (1-f_{\rm d})\,\alpha_{\rm s} + f_{\rm d}\,\alpha_{\rm d}, 
\qquad
\varepsilon_{\rm eff} \approx (1-f_{\rm d})\,\varepsilon_{\rm s} + f_{\rm d}\,\varepsilon_{\rm d}.
\label{eq:alph_eps_mix}
\end{equation}
Radiator capacity scales as $Q \propto \varepsilon_{\rm eff}\,\sigma A\,(T^{4}-T_{\rm sink}^{4})$; thus even modest $f_{\rm d}$ reduces margin if $\varepsilon_{\rm d}<\varepsilon_{\rm s}$.  For solar arrays at low Sun angles, the effective transmittance/reflectance change can be linearized as $\Delta P/P \simeq -\kappa\,f_{\rm d}$ with $\kappa\sim\mathcal{O}(1)$, tuned to array coatings. 

A thin dust layer of thickness $t$ on an RF window (wavelength $\lambda$) and complex permittivity $\varepsilon' (1-j\tan\delta)$ yields one--way power attenuation 
\begin{equation}
L_{\rm dB} \approx 8.686\,\alpha\,t,\quad
\alpha \simeq k_0\,\operatorname{Im}\!\left\{\sqrt{\varepsilon'(1 - j\tan\delta)}\right\},\quad k_0=\frac{2\pi}{\lambda},
\label{eq:rf_loss}
\end{equation}
with $\varepsilon'$ and $\tan\delta$ from Sec.~\ref{sec:elec-dielec-prop} (and Table~\ref{tab:regional}) at the relevant temperature. 
Engineering impacts cited above (albedo reduction, backscatter increase, CF shift) explain observed degradations in radiometry and power (cf. Sec.~\ref{sec:near-surface-0-3m} for near--surface dust densities). 

\paragraph{Thin dust films on optics.}
For a gray dust film of areal coverage $f_{\rm d}$, the effective normal reflectance $R_{\rm eff}$ of a coated surface with clean reflectance $R_0$ 
and dust single-scattering albedo $w$ obeys, to first order in $f_{\rm d}$,
\begin{equation}
R_{\rm eff}\!\simeq\!R_0(1-f_{\rm d}) + \frac{w}{\pi}\,f_{\rm d},
\label{eq:K_uaa}
\end{equation}
so that a few percent coverage ($f_{\rm d}\!\sim\!0.05$) on a dark substrate (e.g., solar cells) reduces $R_{\rm eff}$ roughly linearly with $f_{\rm d}$ 
while increasing backscatter (larger $B_0$, smaller $h$). 

\paragraph{Thermal IR contrast under dust coatings.}
A porous dust overburden of thickness $t$ and conductivity $k$ on a radiator with emissivity $\epsilon_{\rm r}$ adds an areal thermal resistance 
$R_{\rm th} = t/k$, shifting the effective radiative temperature by $\Delta T \simeq q\,R_{\rm th}$ at heat flux $q$;  for $t = 100~\mu$m and $k = 10^{-3}\,{\rm W\,m^{-1}\,K^{-1}}$ (Table~\ref{tab:properties}), $R_{\rm th}\!\approx\!100~{\rm K\,m^2\,W^{-1}}$, which is non-negligible for $q \gtrsim 10~{\rm W\,m^{-2}}$.

\section{Mobilization and transport}
\label{sec:mobilization-transport}

Dust transport operates through three distinct channels with weak coupling except during transients: (i) meteoroid ejecta maintain a persistent, dawn-enhanced exospheric population and mix the upper centimeters on Myr timescales; (ii) electrostatic hopping is intermittent and localized near strong $E$-field gradients at illumination and plasma transitions; (iii) rocket plumes drive brief, intense events with the highest grain velocities $v_{\mathrm{ejecta}} \sim 10^{2}$--$10^{3}~\mathrm{m\,s^{-1}}$. 

In this Section we will quantify each channel; the integrated risk picture maps $(\mathrm{PSD},\rho,n,\varepsilon',k)$ into flux, spectrum, and angular distributions at a site.

\subsection{Meteoroid-generated dust cloud}
\label{sec:meteoroid-dust}

{LADEE}/{LDEX} resolved a permanent, asymmetric dust cloud with dawn enhancement; characteristic detected radii $a\gtrsim 0.3$--0.7 $\mu$m and speeds of order $10^{2}$ m s$^{-1}$; instantaneous mass $\sim 10^{2}$ kg \citep{Horanyi2015}. Impact gardening mixes the upper centimeters on Myr timescales.
Stream encounters produce large, time-dependent enhancements; during the 2013 Geminids, LDEX measured over $10\times$ the number of dense ejecta plumes relative to non-shower periods, constraining source variability and ballistic fluxes \cite{Szalay2018Geminids}.

\subsection{Electrostatic lofting near the surface}

Terminator and shadow transitions impose sharp horizontal and vertical electric-field gradients. Under these conditions, submicron grains can execute short hops from centimeters to meters.  Apollo-era in situ dust sensors provide additional but still debated constraints: the Apollo~17 Lunar Ejecta and Meteorites (LEAM) experiment reported an unexpected enhancement of low-altitude dust impacts near the terminator, though later analyses suggested possible thermal/mechanical artifacts; nevertheless the LEAM timing and directionality remain a useful bound on any terminator-linked near-surface dust component \cite{OBrien2011DustMovements,Grun2013LEAM}.
Surveyor's horizon glow  \cite{RennilsonCriswell1974} is consistent with forward scattering by near-surface grains, while LADEE placed strict upper limits on any dense, sustained, high-altitude $0.1~\mu$m layers \cite{Horanyi2015,Colwell2007,Stubbs2006}. The net picture is intermittent, localized mobilization governed by sheath dynamics rather than a persistent, optically thick layer.

\subsection{Near-surface dust (0--3 m above the surface)}
\label{sec:near-surface-0-3m}

We quantify trajectories of \emph{pre-liberated} grains in the first meters above the surface, where $E(z)$ varies on decimeter scales and the adhesion threshold (Fig.~\ref{fig:E_vs_a_required})  can prevent injection into the sheath  [cf.~Eq.~(\ref{eq:E-adh})]. We parametrize the near-surface sheath by its potential drop
\begin{equation}
\Delta V_{\rm sh} \equiv \int_{0}^{\infty} E(z)\,dz,
\quad {\rm with} \quad
E(z) = E_{0}\exp\!\left(-\frac{z}{\lambda}\right)
\;\;\Rightarrow\;\;
\Delta V_{\rm sh}\simeq E_{0}\lambda,
\label{eq:E-integral} 
\end{equation}
with $E_{0}$ a near-surface amplitude and $\lambda$ the e-folding scale ($\sim\,$3--30~cm near illuminated terminators and shadow edges; Table~\ref{tab:E0lambda}). The terrain--specific envelope implied by Eqs.~(\ref{eq:h_ball})--(\ref{eq:hover}) is summarized in Table~\ref{tab:0to3m} using the field priors of Table~\ref{tab:E0lambda} (Sec.~\ref{sec:elec-dielec-prop}B) and the detachment threshold in Fig.~\ref{fig:E_vs_a_required}.

\begin{table}[htbp]
\centering
\caption[Recommended near-surface field envelopes]{Recommended near-surface field envelopes by terrain/geometry. $E_0$ is a characteristic near-surface amplitude; $\lambda$ is an e-folding scale (use with Eqs.~(\ref{eq:h_ball})--(\ref{eq:hover})).}
\label{tab:E0lambda}
\setlength{\tabcolsep}{2pt}
\renewcommand{\arraystretch}{1.0}

\begin{tabular}{llll}
\hline
\parbox[t]{3.8cm}{\raggedright Setting} &
\parbox[t]{1.8cm}{\raggedright $E_0$ [V\,m$^{-1}$]} &
\parbox[t]{1.5cm}{\raggedright $\lambda$ [m]} &
\parbox[t]{7.9cm}{\raggedright Notes} \\
\hline
\parbox[t]{3.8cm}{\raggedright Terminator / shadow edge} &
\parbox[t]{1.8cm}{\raggedright $10^{2}$--$10^{3}$} &
\parbox[t]{1.5cm}{\raggedright 0.03--0.30} &
\parbox[t]{7.9cm}{\raggedright Strong gradients over decimeters; patchy in space/time.} \\

\parbox[t]{3.8cm}{\raggedright PSR rims / cold traps} &
\parbox[t]{1.8cm}{\raggedright $10^{2}$--$10^{3}$} &
\parbox[t]{1.5cm}{\raggedright 0.03--0.30} &
\parbox[t]{7.9cm}{\raggedright Similar $E_0$; longer $\tau_{\mathrm{relax}}$ sustains patches.} \\

\parbox[t]{3.8cm}{\raggedright Magnetic swirl boundaries} &
\parbox[t]{1.8cm}{\raggedright $10^{2}$--$10^{3}$} &
\parbox[t]{1.5cm}{\raggedright 0.03--0.30} &
\parbox[t]{7.9cm}{\raggedright Modified photoemission/charging; site-specific $E(z)$.} \\

\parbox[t]{3.8cm}{\raggedright Sunlit interiors} &
\parbox[t]{1.8cm}{\raggedright $\le 10^{2}$} &
\parbox[t]{1.5cm}{\raggedright $\le 0.05$} &
\parbox[t]{7.9cm}{\raggedright Weak vertical gradients; intermittent hopping only.} \\
\hline
\end{tabular}
\end{table}

In vacuum the field does work $q\,\Delta V_{\rm sh}$ on a grain of charge $q=4\pi\varepsilon_{0}a\phi_{g}$. Neglecting drag and assuming $q$ remains approximately constant during a short hop (charge-relaxation times $\tau_{\rm relax}$ in cold, porous soils are typically $\gg$ hop times; see Sec.~\ref{sec:dielec-losst}), the ballistic apex height is
\begin{equation}
h_{\rm ball}(a)\;\simeq\;\frac{q\,\Delta V_{\rm sh}}{m\,g_{\rm Moon}}
\;=\;\frac{3\,\varepsilon_{0}\,\phi_{g}}{a^{2}\rho_{p}\,g_{\rm Moon}}\;\Delta V_{\rm sh},
\label{eq:h_ball}
\end{equation}
where $m=\tfrac{4}{3}\pi a^{3}\rho_{p}$. For an exponential sheath, $E(z)=E_{0}e^{-z/\lambda}$, the static ``hover'' height $h_{*}$ (where the upward electrostatic force equals weight) solves $q\,E(h_*) = m g_{\rm Moon}$:
\begin{equation}
h_{*}(a)\;\simeq\;\lambda\,\ln\!\left(\frac{qE_{0}}{m g_{\rm Moon}}\right)
\;=\;\lambda\,\ln\!\left(\frac{3\varepsilon_{0}\phi_{g}E_{0}}{a^{2}\rho_{p}g_{\rm Moon}}\right),
\label{eq:hover}
\end{equation}
valid when $qE_{0}>m g_{\rm Moon}$. Eq.~(\ref{eq:h_ball}) applies once adhesion is already overcome [cf. Eq.~\eqref{eq:E-adh}]; the field needed for detachment is given by Eq.~\eqref{eq:E-adh}, which is dominated by the $F_{\rm adh}$ term across the micron range (Fig.~\ref{fig:E_vs_a_required}). The terrain-specific envelope implied by Eqs.~(\ref{eq:h_ball})--(\ref{eq:hover})  is summarized in Table~\ref{tab:0to3m}. In fact, these heights apply once detachment is achieved (adhesion-aware requirement in Eq.~\eqref{eq:E-adh} and Fig.~\ref{fig:E_vs_a_required}); detachment may be provided by mechanical pre-liberation \emph{and/or} by bed-mediated microcavity charge pumping in the patched-charge model \citep{Wang2016DustChargingTransport,Schwan2017ChargeState,Hood2018RateLofting}.

\paragraph{Including finite launch velocity from bed-mediated ejection.}
Equation~(\ref{eq:h_ball}) assumes negligible initial vertical kinetic energy at detachment. If a grain is ejected from the bed with an initial vertical speed $v_{0}$ (as observed in microcavity/PCM laboratory experiments \citep{Wang2016DustChargingTransport,Carroll2020LaunchVelocities}), the apex height generalizes to
\begin{equation}
h_{\rm ball}(a)\;\simeq\;\frac{q\,\Delta V_{\rm sh}+\tfrac{1}{2}m v_{0}^{2}}{m\,g_{\rm Moon}}
\;=\;\frac{q\,\Delta V_{\rm sh}}{m g_{\rm Moon}}+\frac{v_{0}^{2}}{2g_{\rm Moon}}.
\label{eq:h_ball_v0}
\end{equation}
For example, $v_{0}\sim 0.6~{\rm m\,s^{-1}}$ corresponds to $h\sim v_{0}^{2}/(2g_{\rm Moon})\approx 0.11$~m, consistent with centimeter-scale lofting observed in vacuum-chamber conditions \citep{Wang2016DustChargingTransport}.

\begin{table}[h]
\caption{Expected apex/hover heights for \emph{pre-liberated} grains in the first \SI{0}{-}\SI{3}{\meter} from Eqs.~(\ref{eq:h_ball})--(\ref{eq:hover}) using $E_0,\lambda$ from Table~\ref{tab:E0lambda} and baseline $\phi_g=\SI{5}{\volt}$, $\rho_p=\SI{3.1e3}{\kilogram\per\meter\cubed}$, adhesion satisfied per Eq.~(\ref{eq:E-adh}). Field envelopes $E_{0}$ and e-folds $\lambda$ follow Sec.~\ref{sec:charg-sheath-lift}; hover heights $h_{*}$ use Eq.~(\ref{eq:hover}) with $\phi_{g}=5$~V, $\rho_{p}=3.1\times10^{3}$~kg\,m$^{-3}$. Where ranges are given, $E_{0}=300$~V\,m$^{-1}$, $\lambda=0.03$--$0.30$~m are assumed for the $h_{*}$ examples. These ranges operationalize adhesion--aware transport in the near--surface meter--scale regime.}
\label{tab:0to3m}
\setlength{\tabcolsep}{4pt}
\renewcommand{\arraystretch}{1.12}

\begin{tabular}{llll}
\hline
\parbox[t]{3.2cm}{\raggedright Terrain / setting} &
\parbox[t]{3.0cm}{\raggedright Field envelope (Sec.~\ref{sec:charg-sheath-lift})} &
\parbox[t]{4.4cm}{\raggedright Example $h_{*}$ (pre-liberated)\textsuperscript{a}} &
\parbox[t]{6.0cm}{\raggedright Notes / constraints} \\
\hline\hline

\parbox[t]{3.2cm}{\raggedright Terminator \& sharp shadow edges} &
\parbox[t]{3.0cm}{\raggedright
$E_{0}\!\sim\!10^{2}$--$10^{3}$ \si{\volt\per\meter};\\
$\lambda\!\sim\!$ \SIrange{0.03}{0.30}{\meter}} &
\parbox[t]{4.4cm}{\raggedright
$a=1.0~\mu$m: $0.06$--$0.62$ m;
$a=0.5~\mu$m: $0.10$--$1.04$ m} &
\parbox[t]{6.0cm}{\raggedright
Patchy, time-variable $E(z)$; short hops/hovers within \SIrange{0}{3}{\meter},
(\ref{eq:h_ball})--(\ref{eq:hover}), Fig.~\ref{fig:E_vs_a_required}.}
\\[1.0ex]

\parbox[t]{3.2cm}{\raggedright PSR rims / cold traps} &
\parbox[t]{3.0cm}{\raggedright Similar $E_{0}$; longer $\tau_{\rm relax}$} &
\parbox[t]{4.4cm}{\raggedright Similar $h_{*}$; longer-lived charge patches} &
\parbox[t]{6.0cm}{\raggedright
Low $T$ and high porosity increase $\tau_{\rm relax}$ (Sec.~\ref{sec:dielec-losst});
persistent gradients likely.}
\\[1.0ex]

\parbox[t]{3.2cm}{\raggedright Magnetic swirl boundaries} &
\parbox[t]{3.0cm}{\raggedright Lateral $E$-gradients; modified photoemission} &
\parbox[t]{4.4cm}{\raggedright Comparable $h_{*}$ when $E_{0}$ is in the same band} &
\parbox[t]{6.0cm}{\raggedright
Reduced weathering \& altered charging; grain sorting possible; use site $E(z)$ where available.}
\\[1.0ex]

\parbox[t]{3.2cm}{\raggedright Sunlit interiors (away from boundaries)} &
\parbox[t]{3.0cm}{\raggedright $E_{0}$ smaller, $\lambda$ short} &
\parbox[t]{4.4cm}{\raggedright $h_{*}\lesssim$ few cm--dm} &
\parbox[t]{6.0cm}{\raggedright Weak vertical gradients; intermittent hopping only.}
\\[1.0ex]

\parbox[t]{3.0cm}{\raggedright Meteoroid ejecta (all terrains)} &
\parbox[t]{3.0cm}{\raggedright --} &
\parbox[t]{4.4cm}{\raggedright Broad ballistic arcs; many cm--m cross the 0--3 m layer} &
\parbox[t]{6.0cm}{\raggedright Maintains exosphere; stream spikes (Sec.~\ref{sec:meteoroid-dust}); independent of $E(z)$.}
\\
\hline
\end{tabular}
\vspace{0.5ex}
\footnotesize\raggedright \textsuperscript{a}\,Computed from Eq.~(\ref{eq:hover}) with $\phi_{g}=5$~V,
$\rho_{p}=3.1\times10^{3}$~kg\,m$^{-3}$; $E_{0}$, $\lambda$ from Sec.~\ref{sec:charg-sheath-lift}. \normalsize
\end{table}

\paragraph{Representative  (terminator/shadow-edge) envelope:}
Using $E_{0}=10^{2}$--$10^{3}$~V\,m$^{-1}$ and $\lambda=0.03$--$0.30$~m (Sec.~\ref{sec:charg-sheath-lift}), $\phi_{g}=5$~V, $\rho_{p}=3.1\times10^{3}$~kg\,m$^{-3}$:
\begin{itemize}\setlength{\itemsep}{2pt}
\item $a=1.0~\mu$m: $h_{*}\approx 0.06$--$0.62$~m for $\lambda=0.03$--$0.30$~m at $E_{0}=300$~V\,m$^{-1}$; extending to $\sim\,$2~m for $\lambda\!=\!1$~m (upper bound used in design trades).
\item $a=0.5~\mu$m: $h_{*}\approx 0.10$--$1.04$~m (same envelope).
\item $a=0.1~\mu$m: $h_{*}$ readily exceeds meters if $q$ is maintained. However, LADEE/LDEX placed strict upper limits on any dense, sustained, high-altitude $0.1~\mu$m layers, so persistent populations above $\mathcal{O}(10)$~m are unlikely except during transients (meteoroid streams) or strong local anomalies (Secs.~\ref{sec:meteoroid-dust},\ref{sec:near-surface-0-3m}). 
\end{itemize}

For $a=1~\mu$m, $E_{0}=300$~V\,m$^{-1}$, $\lambda=0.10$~m $\Rightarrow \Delta V_{\rm sh}=30$~V. Then $h_{\rm ball}\!\approx\!0.79$~m and a symmetric hop has $t_{\rm flight}\!\approx\!2\sqrt{2h_{\rm ball}/g_{\rm Moon}}\!\sim\!2$~s, well below typical $\tau_{\rm relax}$ values inferred for cold, porous soils (Sec.~\ref{sec:dielec-losst}), so treating $q$ as quasi-constant during a hop is appropriate. Field envelopes and adhesion thresholds used here are those compiled in Sec.~\ref{sec:charg-sheath-lift}, see Eqs.~(\ref{eq:h_ball})--(\ref{eq:hover}), Fig.~\ref{fig:E_vs_a_required}.  Table~\ref{tab:0to3m} summarizes expected heights by terrain, using site-resolved $E(z)$ where available.

\subsection{Rocket plume entrainment}
\label{sec:rock-plume}

Gas dynamic pressure and wall shear accelerate fines; empirical scalings suggest $\dot{m}_{e}\propto q^{n}$ with $n\approx 2$--3 \citep{Metzger2011}. Apollo 12 dusted Surveyor III at $\sim 155$ m, implying mm--sub-mm ejecta at several hundred m s$^{-1}$ \citep{Carroll1971}. Artemis landers require plume--surface monitors to bound flux, PSD, and angular distributions \citep{Metzger2024}.

Here we use $q$ as the surface energy flux density from the impinging jet at the soil, and $\tau_{w}$ as the wall shear stress. Empirical fits in PSI testing may be written as $\dot{m}_{e}\propto q^{n}$ with $n\simeq 2$--$3$, or alternatively $\dot{m}_{e}\propto \tau_{w}^{m}$ with $m\simeq 1$--$2$, with calibration constants determined per engine and soil state.

Recent re-benchmarking indicates that total eroded mass during Apollo landings was $\sim 4$--$10\times$ larger than earlier estimates, which expands the downrange dust-loading envelope and sandblasting hazard \cite{Metzger2024Part2}. Risk parameterization should therefore be expressed in terms of $\left(q,\;\partial q/\partial r,\;\theta_{\mathrm{plume}}\right)$ and local soil state $(\mathrm{PSD},\rho,n)$, with validation against SCALPSS-class stereo and ejecta-count data on upcoming CLPS and Artemis flights. 

Contemporary PSI instrumentation is maturing: SCALPSS stereo photogrammetry flew on IM-1 (2024) and is planned for additional CLPS and Artemis landings; despite anomalies on IM-1, the campaign yielded engineering lessons and partial data products \cite{Cuesta2025PSI}. New payloads (e.g., EJECTA, DERT, PIE) will constrain ejecta flux, velocity, and PSD in situ to validate PSI models.

Taken together, the updated erosion-rate calibration, event geometry, and SCALPSS-class measurements imply that PSI risk envelopes should be parameterized in terms of $(q,\ \partial q/\partial r,\ \theta_{\rm plume})$, local soil state (PSD, $\rho$, $n$), and landing geometry, with verification against stereo and ejecta-counter data on future CLPS and Artemis flights.

\emph{Operational impacts and mitigations:} Angular fines and elevated specific surface area enhance abrasion, fouling of seals and radiators, and optical scattering. Adhesion forces $F_{\rm adh}\simeq \tfrac{3}{2}\pi W R$ (JKR, with $W\!\sim\!0.1~\mathrm{J\,m^{-2}}$, $R\!\sim\!5~\mu$m) typically exceed lunar gravity for micron-scale grains: $F_{\rm adh}\!\approx\!2.36~\mu\mathrm{N}$ vs.\ $m g_{\rm Moon}\!\approx\!2.10\times10^{-14}~\mathrm{N}$ for $a\!=\!1~\mu$m. Consequently, pre-liberation by vibration, micro-impacts, or gas shear is usually required before sheath acceleration is effective. To reduce risk, (i) minimize dust-entraining geometries near high energy fluxes, (ii) use abrasion-tolerant coatings and sacrificial layers on high-flux surfaces, (iii) design dust-tolerant seals and connectors, (iv) provide localized mechanical agitation prior to electrostatic removal, and (v) co-locate plume/dust monitoring with E-field probes during landings to bound $\dot{m}_{\rm e}$, PSD, and angular distributions for future designs.

\subsection{Mechanisms, observables, and measurements}

\begin{figure}[h]
\centering
\begin{tikzpicture}
\begin{axis}[
  width=0.68\linewidth,
  height=0.38\linewidth,
  xlabel={Time relative to stream peak (hours)},
  ylabel={Dust plume count ratio (to baseline)},
  xmin=-36, xmax=36, ymin=0.5, ymax=12,
  grid=both, ymode=log,
  tick label style={/pgf/number format/fixed},
]
% a schematic bump peaking at 10x
\addplot+[samples=140,domain=-36:36, mark=*, mark size=1.1pt,
          mark options={line width=0.2pt}]
  { 1 + 9*exp(-(x^2)/(2*9)) };
\end{axis}
\end{tikzpicture}
\caption{Schematic of meteoroid-stream forcing (e.g., Geminids, \cite{Szalay2018Geminids}) and the resulting transient enhancements in near-surface dust flux. The persistent background is maintained by sporadics; stream spikes modulate injection independent of $E(z)$ (cf. Sec.~\ref{sec:near-surface-0-3m}).  Stream forcing sets event-driven upper bounds on exospheric dust.}
\label{fig:stream}
\end{figure}
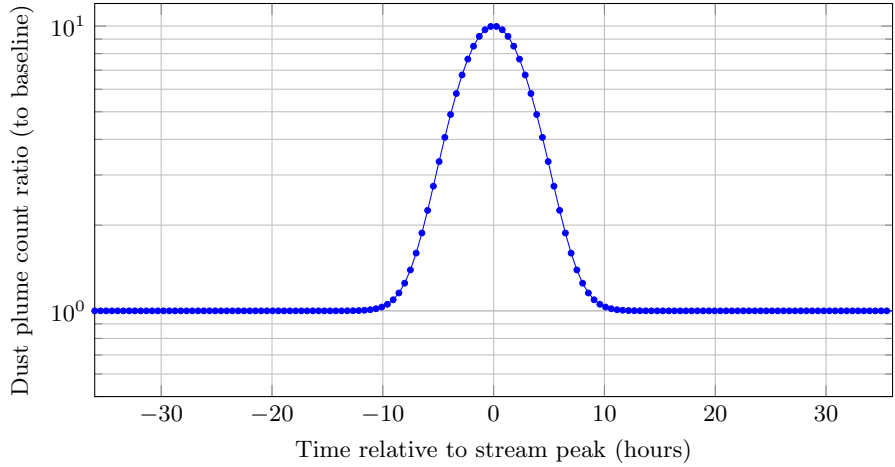

The three mobilization channels leave distinct, jointly observable signatures. Table~\ref{tab:mechanism-observables} cross-walks mechanisms to observables and heritage constraints. Meteoroid ejecta maintain a persistent, asymmetric dust cloud with stream-driven enhancements (Fig.~\ref{fig:stream}); electrostatic hopping is intermittent and localized near strong sheath gradients at illumination or plasma boundaries; rocket-plume entrainment produces brief, intense episodes with the highest $v_{\rm ejecta}$ and the greatest engineering consequences. Co-registered electric fields, grain charge, PSD, and flux near terminators, PSR rims, and landing sites resolve model degeneracies and tighten design ranges by directly linking source, transport, and redeposition.

\begin{table*}[h]
\centering
\caption{Cross-walk from production/modification mechanisms to primary observables/key constraints used in this review.}
\label{tab:mechanism-observables}
\setlength{\tabcolsep}{2pt}
\renewcommand{\arraystretch}{1.0}

\begin{tabular}{lll}
\hline
\parbox[t]{0.28\textwidth}{\raggedright Mechanism} &
\parbox[t]{0.35\textwidth}{\raggedright Primary observables} &
\parbox[t]{0.33\textwidth}{\raggedright Key missions / instruments} \\
\hline\hline

\parbox[t]{0.28\textwidth}{\raggedright Impact comminution, agglutination} &
\parbox[t]{0.35\textwidth}{\raggedright PSD shape; agglutinate fraction; maturity indices (Is/FeO)} &
\parbox[t]{0.33\textwidth}{\raggedright Apollo samples (imaging, sieving), \cite{Carrier1991,Carrier2003,Morris1978}} \\[0.6ex]

\parbox[t]{0.28\textwidth}{\raggedright Space weathering (npFe$^{0}$)} &
\parbox[t]{0.35\textwidth}{\raggedright Albedo reduction; CF shift; suppression of mafic band depths} &
\parbox[t]{0.33\textwidth}{\raggedright LRO/Diviner CF; M$^{3}$ (Chandrayaan-1), \cite{Lucey2017CF,Green2011}} \\

\parbox[t]{0.28\textwidth}{\raggedright Thermal fatigue} &
\parbox[t]{0.35\textwidth}{\raggedright Boulder-size decay; fines production vs age} &
\parbox[t]{0.33\textwidth}{\raggedright Diviner rock abundance vs crater age \cite{Bandfield2011}} \\[0.6ex]

\parbox[t]{0.28\textwidth}{\raggedright Pyroclastic fragmentation} &
\parbox[t]{0.35\textwidth}{\raggedright Glass-rich spectra; CF diagnostics} &
\parbox[t]{0.33\textwidth}{\raggedright Diviner mapping of pyroclastic mantling deposits, \cite{Allen2012}} \\

\parbox[t]{0.28\textwidth}{\raggedright Near-surface charging} &
\parbox[t]{0.35\textwidth}{\raggedright Sheath electric field; lofting thresholds; negative ions} &
\parbox[t]{0.33\textwidth}{\raggedright Surveyor horizon glow; LADEE constraints; CE-6 NILS H$^{-}$, \cite{Colwell2007,Stubbs2006,Wieser2025NILSHminus}} \\

\parbox[t]{0.28\textwidth}{\raggedright Exospheric maintenance} &
\parbox[t]{0.35\textwidth}{\raggedright Dust flux vs local time and meteoroid streams} &
\parbox[t]{0.33\textwidth}{\raggedright LADEE/LDEX cloud and stream responses, \cite{Horanyi2015,Szalay2018Geminids}} \\

\parbox[t]{0.28\textwidth}{\raggedright Plume--surface entrainment (PSI)} &
\parbox[t]{0.35\textwidth}{\raggedright $v_{\mathrm{ejecta}}\sim 10^{2}$--$10^{3}$ m\,s$^{-1}$; $\dot{m}_{e}\propto q^{n}$, $n\approx 2$--$3$; erosion scales with engine energy flux} &
\parbox[t]{0.33\textwidth}{\raggedright Surveyor III coating; Apollo; SCALPSS and PSI payloads, \cite{Carroll1971,Metzger2011,Metzger2024,Cuesta2025PSI}} \\
\hline
\end{tabular}
\end{table*}

\section{Regional reference values (design ranges)}
\label{sec:reg_refvalues}

\subsection{Regional narratives} 
Maria, highlands, swirls, pyroclastics, and polar PSRs are end-members in a continuum. Maria typically combine higher FeO and variable TiO$_{2}$ with lower albedo and, locally, higher rock abundance; their dielectric response tends to be higher with ilmenite content and compaction. Highlands are plagioclase rich with lower FeO, brighter albedo, thicker regolith, and a dielectric response dominated by lower-density silicates. Swirls sit atop magnetic anomalies that partially shield solar wind, reducing space weathering rates and producing optically immature lanes; electrostatic grain sorting may further separate fines. Pyroclastics include glass-rich mantling units with distinct CF positions and mid-IR signatures. PSR floors push toward extreme $T$, slow charge relaxation, and potential frost cements; rims and near-PSR slopes mix cold-regime charging with occasional direct illumination and enhanced horizontal $E$-fields. The row added for Vikram lander's site highlights how in-situ thermophysics at high southern latitude modifies expectations for $k$ and $\rho$ at centimeter depths. Use these as priors and propagate to thermal and electrical behavior with Eqs.~\eqref{eq:rhoz} and \eqref{eq:k-twochannel} and the Looyenga mixing in Eq.~\eqref{eq:Looyenga}.

\subsection{Design takeaways by region} 

Maria: higher FeO/TiO$_{2}$ and compaction bias toward higher $\varepsilon'$ and $k$, increasing RF losses and thermal conduction. Highlands: lower FeO and thicker regolith reduce $\varepsilon'$, with lower near-surface $k$. Swirls: reduced space weathering maintains high albedo and may alter charge balance along anomaly boundaries. Pyroclastics: glass-rich fines shift CF and may reduce grain strength. PSR floors: extreme $T$ and long charge-relaxation times increase the likelihood of persistent fields; frost cements alter mechanics. Vikram site: in-situ ChaSTE shows $k$ at 80 mm that is several times the Diviner near-surface reference, emphasizing centimeter-scale structure.  

\begin{table}[h]
\centering
\caption{Regional reference values (upper meters unless noted). Use as design references and adjust with site-resolved data. Dielectric values are referenced to LPR frequencies (decimeter--meter wavelengths) and near-surface temperatures.}
\label{tab:regional}
\setlength{\tabcolsep}{2pt}
\renewcommand{\arraystretch}{1.0}

\begin{tabular}{lllll}
\hline
\parbox[t]{1.7cm}{\raggedright Region} &
\parbox[t]{1.5cm}{\raggedright PSD me- dian ($\mu$m)} &
\parbox[t]{2.75cm}{\raggedright Bulk density $\rho$ at 0\,m / 1\,m (g cm$^{-3}$)} &
\parbox[t]{3.6cm}{\raggedright $\varepsilon'$ / $\tan\delta$ (upper meters)} &
\parbox[t]{6.8cm}{\raggedright Notes} \\
\hline\hline

\parbox[t]{1.7cm}{\raggedright Maria} &
\parbox[t]{1.5cm}{\raggedright 50--120} &
\parbox[t]{2.75cm}{\raggedright 1.4 / 1.85} &
\parbox[t]{3.6cm}{\raggedright 3.0--3.5 / $10^{-3}$--$10^{-2}$} &
\parbox[t]{6.8cm}{\raggedright Elevated FeO, TiO$_{2}$; low albedo; higher rock fraction \cite{Papike1998,Bandfield2011}.} \\

\parbox[t]{1.7cm}{\raggedright Highlands} &
\parbox[t]{1.5cm}{\raggedright 60--140} &
\parbox[t]{2.75cm}{\raggedright 1.3 / 1.80} &
\parbox[t]{3.6cm}{\raggedright 2.6--3.1 / $10^{-3}$--$10^{-2}$} &
\parbox[t]{6.8cm}{\raggedright Plagioclase rich; brighter; thicker regolith \cite{Papike1998}.} \\

\parbox[t]{1.7cm}{\raggedright Pyroclastics} &
\parbox[t]{1.5cm}{\raggedright 40--100} &
\parbox[t]{2.75cm}{\raggedright 1.4 / 1.85} &
\parbox[t]{3.6cm}{\raggedright 3.1--3.6 / $10^{-3}$--$10^{-2}$} &
\parbox[t]{6.8cm}{\raggedright Glass-rich; distinct CF; Fe/Ti often high \cite{Allen2012}.} \\

\parbox[t]{1.7cm}{\raggedright Swirls} &
\parbox[t]{1.5cm}{\raggedright 50--120} &
\parbox[t]{2.75cm}{\raggedright 1.4 / 1.85} &
\parbox[t]{3.6cm}{\raggedright 2.8--3.3 / $10^{-3}$--$10^{-2}$} &
\parbox[t]{6.8cm}{\raggedright Reduced weathering; optically immature lanes \cite{Blewett2011,Glotch2015}.} \\

\parbox[t]{1.7cm}{\raggedright PSR floors} &
\parbox[t]{1.5cm}{\raggedright 60--150} &
\parbox[t]{2.75cm}{\raggedright 1.35 / 1.80} &
\parbox[t]{3.6cm}{\raggedright 2.5--3.2 / $10^{-3}$--$10^{-2}$} &
\parbox[t]{6.8cm}{\raggedright 25--40 K; negative charging; possible frost cements \cite{Paige2010,Hayne2017}.} \\[0.6ex]

\parbox[t]{1.7cm}{\raggedright Vikram site\\(69.373${}^\circ$~S,\,32.319${}^\circ$~E)} &
\parbox[t]{1.5cm}{\raggedright 40--120} &
\parbox[t]{2.75cm}{\raggedright 1.94 / --} &
\parbox[t]{3.6cm}{\raggedright {not measured} / {not measured}} &
\parbox[t]{6.8cm}{\raggedright ChaSTE at 80 mm: $k \approx 1.2\times 10^{-2}$\,W\,m$^{-1}$\,K$^{-1}$; local compaction \cite{Mathew2025ChaSTE}.} \\
\hline
\end{tabular}
\vspace{2pt}
\begin{minipage}{0.98\linewidth}
\footnotesize
\justifying
\textit{Provenance and uncertainty.} The ranges in Table~\ref{tab:regional} are intended as engineering priors that bracket both geographic variability and inversion/model systematics. PSD medians are anchored by returned-sample distributions but are sensitive to comminution state and electrostatic lofting; treat as $\sim\times 2$ uncertainty in median radius at fixed location. Bulk densities/porosities are based on Apollo core tubes and Diviner thermophysical inversions; use $\pm 0.1$--$0.2$~g\,cm$^{-3}$ as a typical site-level uncertainty. Dielectric parameters are referenced to decimeter--meter wavelengths (LPR-scale) and dry soils; systematic errors from radar inversion and temperature dependence can reach $\mathcal{O}(10\%)$ in $\varepsilon'$ and factors of $\sim 2$ in $\tan\delta$ for low-loss materials. Cooling from 293$\rightarrow$213~K reduces $\varepsilon'$ by $\sim$6--18\% for common end-members at fixed porosity \cite{Kobayashi2023TempPerm}; below $\sim$200~K data remain sparse and should be treated as an additional systematic in PSRs.
\end{minipage}
\end{table}

\paragraph{Design call-out for polar traverses.}
For cold, porous soils near PSR rims, long $\tau_{\rm relax}$ (Sec.~\ref{sec:dielec-losst}) increases the likelihood of persistent charge patches. Combined with high-SSA, angular fines, this elevates \emph{adhesion} and \emph{abrasion} risks to radiators, optics, seals, and electrical connectors. Trafficability envelopes should therefore use the upper-end $(\phi,c)$ in Table~\ref{tab:regional} with frost-cement contingencies 
(PSR floors row in Table~\ref{tab:regional}; frost-cement contingency, Eqs.~(\ref{eq:PSR-eq_6})--(\ref{eq:PSR-eq_7}), and dust-mitigation strategies should assume that electrostatic removal is ineffective unless mechanical pre-liberation is provided (Fig.~\ref{fig:E_vs_a_required}, Table~\ref{tab:adhesion}). Operationally, minimize prolonged loitering at strong illumination/plasma boundaries during high meteoroid-flux windows (Sec.~\ref{sec:meteoroid-dust}), when both ejecta and charging transients are maximized.

\section{Engineering reference ranges (global)}
\label{sec:eng_refvalues}

Table~\ref{tab:regional} summarizes regional reference regolith values, while Table~\ref{tab:properties} provides representative lunar dust/regolith properties. To use these ranges in design, we recommend the following approach:
\begin{enumerate}\setlength{\itemsep}{2pt}
\item Select a regional prior from Table~\ref{tab:regional} (maria, highlands, pyroclastics, swirls, PSR) and latitude.
\item Initialize $\rho(z)$ with $\rho_{0}$, $\rho_{\infty}$, and $z_{c}$; propagate $k(T,\rho)$ to obtain $\Gamma$ and the diurnal $T(z,t)$.
\item Use composition and densification to set $\varepsilon'$ and $\tan\delta$ (Looyenga mixing).
\item Evaluate adhesion and Mohr--Coulomb parameters $(\phi,c)$ for detachment thresholds and trafficability.
\item For plume events, parameterize erosion and $v_{\mathrm{ejecta}}$ by $(q,\ \partial q/\partial r,\ \theta_{\mathrm{plume}})$ and local soil state (PSD, $\rho$, $n$), and verify against SCALPSS-class stereo and ejecta-counter data where available.
\end{enumerate}

\begin{table}[h!]
\centering
\caption{Representative lunar dust/regolith properties; values depend on grain size, maturity, depth, composition,  region.}
\label{tab:properties}
\renewcommand\arraystretch{1.0}
\setlength{\tabcolsep}{2pt}

\begin{tabular}{lll}
\hline
\parbox[t]{3.9cm}{\raggedright Property} &
\parbox[t]{6.3cm}{\raggedright Representative values} &
\parbox[t]{6.9cm}{\raggedright Notes/primary constraints} \\
\hline \hline

\parbox[t]{3.9cm}{\raggedright Particle size distribution (PSD)} &
\parbox[t]{6.3cm}{\raggedright Median $\sim$\,40--130\,$\mu$m; fines $<20\,\mu$m abundant} &
\parbox[t]{6.9cm}{\raggedright Broad, well-graded PSD; agglutinates common, Apollo sieving and imaging \citep{Carrier2003,Carrier1991}.} \\

\parbox[t]{3.9cm}{\raggedright Specific surface area (SSA)} &
\parbox[t]{6.3cm}{\raggedright $\sim$0.02--0.78 m$^{2}$ g$^{-1}$; typical $\sim$0.5~\si{m^2 g^{-1}}} &
\parbox[t]{6.9cm}{\raggedright Irregularity $\sim$8$\times$ vs spheres; BET; angular grains \citep{Carrier1991,Cadenhead1977,Kring2006}.} \\

\parbox[t]{3.9cm}{\raggedright Bulk density $\rho_{\rm bulk}$} &
\parbox[t]{6.3cm}{\raggedright From 1.3--1.6 g cm$^{-3}$ (surface) to $\gtrsim$\,1.7--1.9 g cm$^{-3}$ (1 m)} &
\parbox[t]{6.9cm}{\raggedright Exponential compaction with depth \cite{Carrier1991}.} \\

\parbox[t]{3.9cm}{\raggedright Friction angle $\phi$, cohesion $c$} &
\parbox[t]{6.3cm}{\raggedright $\phi\sim$\,30--50$^{\circ}$ (to $\sim55^{\circ}$ by \SI{1}{\meter}); $c\sim$\,0.1--1~kPa (surface) to $>$3~kPa (\SIrange{0.5}{1}{\meter})} &
\parbox[t]{6.9cm}{\raggedright Apollo/Lunokhod/analogs geotechnics \citep{Carrier1991,Connolly2023}.} \\

\parbox[t]{3.9cm}{\raggedright Thermal conductivity $k$} &
\parbox[t]{6.3cm}{\raggedright $\sim$\,$7.4\times10^{-4}$ (surface) to $\sim$\,$3.4\times10^{-3}$~W\,m$^{-1}$\,K$^{-1}$ (1 m)} &
\parbox[t]{6.9cm}{\raggedright Diviner inversions \citep{Hayne2017}.} \\

\parbox[t]{3.9cm}{\raggedright $k$ (80 mm, 69.373${}^\circ$)} &
\parbox[t]{6.3cm}{\raggedright $(1.15$--$1.24)\times 10^{-2}$ W m$^{-1}$ K$^{-1}$} &
\parbox[t]{6.8cm}{\raggedright ChaSTE in situ \cite{Mathew2025ChaSTE}.} \\

\parbox[t]{3.9cm}{\raggedright PSR $T_\mathrm{surf}$} &
\parbox[t]{6.3cm}{\raggedright $\sim$\SIrange{25}{40}{\kelvin}} &
\parbox[t]{6.9cm}{\raggedright Diviner \citep{Paige2010}.} \\

\parbox[t]{3.9cm}{\raggedright Dielectric $\varepsilon^\prime$, $\tan\delta$} &
\parbox[t]{6.3cm}{\raggedright $\varepsilon^\prime\sim2.5$--$3.5$; $\tan\delta \sim 10^{-3}$--$10^{-2}$ (upper meters)} &
\parbox[t]{6.9cm}{\raggedright CE-4 LPR; ilmenite/density dependence \citep{Dong2021,Giannakis2021}.} \\

\parbox[t]{3.9cm}{\raggedright Maturity indices} &
\parbox[t]{6.3cm}{\raggedright $I_\mathrm{s}$/FeO (FMR), OM/CF trends} &
\parbox[t]{6.9cm}{\raggedright \textit{np}Fe$^{0}$ abundance vs exposure \citep{Morris1978,Lucey2006,Lucey2017CF}.} \\

\parbox[t]{3.9cm}{\raggedright Near-IR hydration} &
\parbox[t]{6.3cm}{\raggedright Variable $2.8$--$3.0\,\mu$m band; diurnal/latitudinal dependence} &
\parbox[t]{6.9cm}{\raggedright M$^{3}$, follow-up \citep{Pieters2009,Bandfield2018}.} \\

\parbox[t]{3.9cm}{\raggedright Exospheric dust} &
\parbox[t]{6.3cm}{\raggedright Persistent, meteoroid-sourced; dawn-enhanced; $a\gtrsim 0.3$--$0.7$ $\mu$m; $v\sim 10^{2}$\,m\,s$^{-1}$} &
\parbox[t]{6.9cm}{\raggedright LADEE/LDEX \cite{Horanyi2015}.} \\

\parbox[t]{3.9cm}{\raggedright Plume--surface entrainment (PSI)} &
\parbox[t]{6.3cm}{\raggedright $v_{\rm ejecta}\sim 10^{2}$--$10^{3}$ m s$^{-1}$; $\dot{m}_{e}\propto q^{n}$, $n\approx 2$--3; erosion scales with engine energy flux} &
\parbox[t]{6.9cm}{\raggedright Surveyor III, Apollo, modern PSI \cite{Carroll1971,Metzger2011,Metzger2024}.} \\
\hline
\end{tabular}
\vspace{2pt}
\begin{minipage}{0.98\linewidth}\footnotesize
\justifying
\textit{Interpretation of ranges.} Unless noted otherwise, values in Table~\ref{tab:properties} are order-of-magnitude envelopes combining inter-site variability and measurement/model systematics. When propagating into the charge-relaxation relation $\tau_{\rm relax}\simeq \varepsilon_{0}\varepsilon'/\sigma$ (Sec.~\ref{sec:dielec-losst}), Eqs.~(\ref{eq:h_ball})--(\ref{eq:hover}), and the RF-loss relation in Eq.~(\ref{eq:rf_loss}), the dominant uncertainties are typically (i) depth- and temperature-dependent conductivity $\sigma$ (hence $\tau_{\rm relax}$), and (ii) frequency/temperature dependence of $\varepsilon'$ and $\tan\delta$. For cold soils, $\tau_{\rm relax}$ may vary by many orders of magnitude with small changes in pore-network connectivity; design trades should therefore treat $\sigma$ (or $\tau_{\rm relax}$) as a log-uncertain parameter and bracket at least $\pm 2$ decades around the nominal values.
\end{minipage}

\end{table}

\section{Model relations useful in design}
\label{sec:design}

The equations below formalize three scale-bridges used throughout. First, microstructure and composition set $\varepsilon'$ and $\tan\delta$ via compositional mixing and densification. Second, the thermophysical profile $k(T,\rho)$ sets the diurnal $T$ wave and thereby the photoelectron current and charge relaxation time, which together determine whether $qE$ can overcome $mg_{\rm Moon}+F_{\rm adh}$. Third, heliophysical forcing (meteoroid streams, solar wind variability) sets the stochastic envelope for ejecta production and sheath variability; where rock abundance and topography focus gas flow or $E$-fields, transport becomes locally efficient. The model relations that follow supply the parameters needed to propagate these effects into design calculations. Table~\ref{tab:design-mapping} provides design mapping from dust/regolith properties to rover subsystems.  

\subsection{Electrostatic lift with adhesion}
\label{sec:model-adhesion}

For a sphere on a flat in the JKR regime, the pull-off force is
$F_{\rm adh}\simeq \tfrac{3}{2}\pi W R$. With $W=0.1~\mathrm{J\,m^{-2}}$ and $R=5~\mu\mathrm{m}$, $F_{\rm adh}=2.36~\mu\mathrm{N}$. For a $1~\mu\mathrm{m}$ grain ($\rho_p=3100~\mathrm{kg\,m^{-3}}$), the weight is $m g_{\rm Moon}=2.10\times 10^{-14}~\mathrm{N}$, so adhesion exceeds gravity by $\sim 1.12\times 10^{8}$. This justifies the need for pre-liberation (vibration, micro-impacts, gas shear) before sheath acceleration. 

\subsection{Dielectric mixing (Looyenga form)}

With silicate matrix $\varepsilon_{s}\approx 2.7$ and ilmenite $\varepsilon_{ilm}\approx 80$ (RF), modest $v_{ilm}=0.02$--0.05 raises $\varepsilon'$ by several tenths when combined with densification, consistent with {LPR} trends.

\subsection{Radiative term in thermal conductivity}
\label{sec:rad-term}

At $T=380$~K, the radiative term contributes $A T^{3}\approx (2\times 10^{-4}$--$2\times 10^{-3})~\mathrm{W\,m^{-1}\,K^{-1}}$ for 
$A \approx (3.6\times 10^{-12}$--$3.6\times 10^{-11})~\mathrm{W\,m^{-1}\,K^{-4}}$, matching Diviner day/night asymmetry.

\begin{table*}[t]
\caption{Design mapping from dust/regolith properties to rover subsystems. 
Variables and equations refer to sections/labels in this paper; use regional priors from Table~\ref{tab:regional} and global ranges in Table~\ref{tab:properties}.}
\label{tab:design-mapping}
\setlength{\tabcolsep}{6pt}
\setlength{\tabcolsep}{2pt}
\renewcommand{\arraystretch}{1.0}
\centering
\begin{tabular}{lll}
\hline
\parbox[t]{3.25cm}{\raggedright Subsystem} &
\parbox[t]{5.00cm}{\raggedright Dust/regolith property} &
\parbox[t]{8.35cm}{\raggedright Design equation / use (pointer)} \\
\hline\hline

\parbox[t]{3.25cm}{\raggedright Mobility (wheels)} &
\parbox[t]{5.00cm}{\raggedright $\phi$, $c$, $\rho(z)$, PSD, $k(T,\rho)$} &
\parbox[t]{8.35cm}{\raggedright Traction/sinkage: Eqs.~(\ref{eq:tau_max_mc})--(\ref{eq:rolling_res}); density vs depth Eq.~(\ref{eq:rhoz}); thermal state via $k(T,\rho)$ Eq.~(\ref{eq:k-twochannel}).} \\

\parbox[t]{3.25cm}{\raggedright Power/thermal (radiators, arrays)} &
\parbox[t]{5.00cm}{\raggedright $\alpha_{\rm d}$, $\varepsilon_{\rm d}$, $f_{\rm d}$; near--surface dust density (0--3 m)} &
\parbox[t]{8.35cm}{\raggedright Coating mix Eq.~(\ref{eq:alph_eps_mix}); near--surface heights Eqs.~(\ref{eq:h_ball})--(\ref{eq:hover}) and Table~\ref{tab:0to3m} for dust exposure.} \\

\parbox[t]{3.25cm}{\raggedright Mechanisms \& seals} &
\parbox[t]{5.00cm}{\raggedright SSA, angularity, adhesion $F_{\rm adh}$, PSD} &
\parbox[t]{8.35cm}{\raggedright Adhesion threshold Eq.~(\ref{eq:E-adh}) and Fig.~\ref{fig:E_vs_a_required}; Mohr--Coulomb parameters (Sec.~\ref{sec:mech-prop}) for intrusion loads.} \\

\parbox[t]{3.25cm}{\raggedright Optics/sensors} &
\parbox[t]{5.00cm}{\raggedright $w$, $B_{0},h$, $\bar\theta$, CF, maturity} &
\parbox[t]{8.35cm}{\raggedright Photometric trends (Sec.~\ref{sec:opt-prop}); engineering impacts paragraph; use regional priors (Table~\ref{tab:regional}).} \\

\parbox[t]{3.25cm}{\raggedright RF windows/antennas} &
\parbox[t]{5.00cm}{\raggedright $\varepsilon'$, $\tan\delta$, $t$, $f$} &
\parbox[t]{8.35cm}{\raggedright Thin--film insertion loss (\ref{eq:rf_loss}); material selection  Sec.~\ref{sec:elec-dielec-prop}.} \\

\parbox[t]{3.25cm}{\raggedright Charging/ESD} &
\parbox[t]{5.00cm}{\raggedright $\varepsilon'$, $\sigma(T,n)$, $E_{0},\lambda$, H$^{-}$} &
\parbox[t]{8.35cm}{\raggedright Relaxation $\tau_{\rm relax}$ (Sec.~\ref{sec:elec-dielec-prop}); lift with adhesion Eqs.~(\ref{eq:E-adh}), (\ref{eq:hover}); NILS update (Sec.~\ref{sec:elec-dielec-prop}).} \\

\parbox[t]{3.25cm}{\raggedright PSI (landing environments)} &
\parbox[t]{5.00cm}{\raggedright PSD, $\rho$, $n$, plume $q,\partial q/\partial r,\theta_{\rm plume}$} &
\parbox[t]{8.35cm}{\raggedright Erosion/flux scalings (Sec.~\ref{sec:rock-plume}); use SCALPSS--class constraints for site priors.} \\
\hline
\end{tabular}
\end{table*}

\subsection{Design implications and measurement priorities}
\label{sec:implications}

The present synthesis reduces unknowns by connecting particle-scale structure and composition to bulk properties and, in turn, to mobilization channels. For mission design, the practical question is not whether dust moves, but when, where, and with what spectrum and flux. The measurements prioritized below close specific loops: they co-register $E$-field, charge, PSD, and flux in the environments where the models are most sensitive (terminators, PSR boundaries, magnetic anomalies, and active landing sites).

Reducing design margins requires: (1) colocated measurements of near-surface $E$-fields, grain charge, and PSD to close the lofting budget; (2) time-resolved dust flux vs meteor streams to quantify ejecta maintenance; (3) PSR microphysics at centimeter depths (dielectric, conductivity, frost fraction); (4) anomaly traverses across swirls to decouple magnetic shielding from electrostatic sorting; (5) standardized in-situ geotechnical profiling with depth; (6) plume--surface interaction monitors to bound $\dot{m}_{e}$, PSD, and angular distributions during landings. (7) Polar-site in-situ thermophysics: replicate ChaSTE-style active heating at multiple depths to tie $k(T,\rho)$ to Diviner retrievals at high latitudes \cite{Mathew2025ChaSTE}.
(8) Near-surface plasma and dust charging: colocate E-field probes with negative-ion and electron spectrometers to quantify sheath composition and charging/lofting budgets in daylight \cite{Wieser2025NILSHminus,CanuBlot2025NILS}.
(9) PSI monitors: fly SCALPSS-class stereo cameras plus ejecta counters on CLPS and Artemis landers to bound $\dot{m}_{e}$, ejecta PSD, and angular distributions during landings \cite{Cuesta2025PSI}.

\paragraph{Dust mitigation and operations: physics-limited requirements.} Eqs.~(\ref{eq:E-adh})--(\ref{eq:C_no-adh}) highlight an important design implication: for $a\lesssim 10~\mu$m, the adhesion term $C_{\rm adh}/a$ typically dominates gravity by many orders of magnitude. Using the nominal contact parameters in Table~\ref{tab:adhesion} ($W=0.1$~J\,m$^{-2}$, $R=5~\mu$m, $\phi_g=5$~V), Eq.~(\ref{eq:E-adh}) gives $E_{\rm adh}(1~\mu\mathrm{m})\approx 4\times10^{9}$~V\,m$^{-1}$ whereas the gravity-only term is $E_{\rm no}(1~\mu\mathrm{m})\approx 40$~V\,m$^{-1}$. Therefore the natural near-surface fields compiled in Table~\ref{tab:E0lambda} ($E_0\sim10^{2}$--$10^{3}$~V\,m$^{-1}$) cannot detach strongly adhered micron-scale dust from engineering surfaces; electrostatics primarily controls transport \emph{after} grains have been mechanically liberated (impacts/abrasion, plume scouring, vibration) or deposited in weak-contact configurations.

In terms of mitigation taxonomy \cite{Abel2023DustGuide,Fritz2024Roadmap}, this motivates two complementary design levers:
\begin{enumerate}
\item \textit{Reduce adhesion and contact activation} (materials/coatings, surface texturing, minimizing dust trapping geometries, controlled vibration, and contamination-tolerant clearances). Because $E_{\rm adh}\propto W R/(\phi_g a)$ via Eq.~(\ref{eq:C_no-adh}), even modest reductions in effective $W$ or asperity radius $R$ translate directly into lower detachment thresholds.
\item \textit{Actively remove or redirect pre-liberated grains} (electrodynamic dust shields, gas or mechanical blow-off, electrostatic deflection, purge flows, filtration/airlocks). For example, an electrodynamic dust shield (EDS) has now been demonstrated on the lunar surface with reported dust-removal fractions of 82\% from a radiator surface and 97\% from glass over multiple cleaning cycles \cite{Buhler2025EDS}, suggesting that repeated-cycle transport (rather than single-shot electrostatic ``lift-off'') is a viable approach for optics/thermal-control surfaces subject to weakly bound deposits.
\end{enumerate}

\paragraph{Worked design example: terminator pre-liberated grain hop.}
Using the terminator/shadow-edge envelope in Sec.~\ref{sec:near-surface-0-3m} with $E_{0}=300$~V\,m$^{-1}$ and $\lambda=0.10$~m, Eq.~(\ref{eq:E-integral}) gives $\Delta V_{\rm sh}\simeq E_{0}\lambda\approx 30$~V. For a $a=1~\mu$m grain at $\phi_g=5$~V, $q=4\pi\varepsilon_{0}a\phi_g\approx 5.6\times10^{-16}$~C and $m=\tfrac{4}{3}\pi a^{3}\rho_p\approx 1.3\times10^{-14}$~kg (with $\rho_p=3.1\times10^{3}$~kg\,m$^{-3}$), so Eq.~(\ref{eq:h_ball}) yields a ballistic apex height $h_{\rm ball}\approx q\Delta V_{\rm sh}/(m g_{\rm Moon})\approx 0.8$~m and a flight time $t_{\rm flight}\sim2\sqrt{2h_{\rm ball}/g_{\rm Moon}}\approx 2$~s. Meanwhile Eq.~(\ref{eq:hover}) gives an equilibrium hover height $h_{*}\approx \lambda\ln\!\left(qE_{0}/(m g_{\rm Moon})\right)\approx 0.21$~m. This example illustrates why (i) realistic sheath voltages can support cm--m motion of \emph{pre-liberated} micron grains, but (ii) the adhesion-limited detachment condition in Eq.~(\ref{eq:E-adh}) must be treated separately in contamination-control trades.

\section{Conclusions}
\label{sec:concl}

Lunar dust emerges from impact comminution, space weathering (npFe$^{0}$ in rims and agglutinates), thermal fatigue, and volcanic fragmentation. The resulting microphysics (PSD, angularity/SSA, agglutinate fraction, composition/oxidation) sets the bulk thermophysical, dielectric, mechanical, and optical response that governs mobilization under local illumination and plasma conditions. Our main conclusions are:

\begin{enumerate}\setlength{\itemsep}{4pt}
\item \textit{Sources $\rightarrow$ properties.} Three microphysical levers dominate: (i) PSD and angularity, (ii) glassy agglutinates with npFe$^{0}$, and (iii) composition/oxidation state. These map to $k(T,\rho)$, $\varepsilon'$, $\tan\delta$, shear strength $(\phi,c)$, and adhesion at asperities (Secs.~\ref{sec:formation}, \ref{sec:phys-chem}).
\item \textit{Thermal and dielectric structure.} A two--channel $k(T,\rho)$ (solid network + pore radiative) reproduces Diviner day/night asymmetry and depth trends; ChaSTE anchors centimeter--depth $k$ at high southern latitude. Densification from $\sim 1.4$ to $\sim 1.9$~g\,cm$^{-3}$ raises $\varepsilon'$ by $\sim$10-30\% (LPR--consistent), and charge--relaxation times span $10^{2}$-$10^{6}$~s across realistic $\sigma$ (Sec.~\ref{sec:thermphys-prop}, \ref{sec:elec-dielec-prop}).
\item \textit{Adhesion--limited lofting.} Adhesion exceeds gravity by $>10^{7}$ for $a\sim 1~\mu$m at representative $W$ and $R$, so pre--liberation (vibration, micro-impacts, gas shear) is typically required. The adhesion--aware field requirement $E_{\rm req}(a) \simeq C_{\rm no}\,a^{2} + C_{\rm adh}/a$ (with $C_{\rm no}$ and $C_{\rm adh}$ given in Eq.~(\ref{eq:C_no-adh})) explains why typical sheath fields ($10^{2}$-$10^{3}$~V\,m$^{-1}$) do not detach adhered grains (Fig.~\ref{fig:E_vs_a_required}, Table~\ref{tab:adhesion}).
\item \textit{Near--surface (0-3 m) transport.} Once detached, micron/sub--micron grains can execute short hops or ``hover'' at heights set by the sheath field integral and scale length; Eqs.~(\ref{eq:h_ball})--(\ref{eq:hover}) bound apex/hover heights and explain the observed first--meters environment near strong gradients (Sec.~\ref{sec:near-surface-0-3m}, Table~\ref{tab:0to3m}).
\item \textit{Mobilization budgets.} Meteoroid ejecta maintain a persistent, dawn--enhanced cloud (with stream--driven spikes), electrostatic hopping is intermittent and localized, and rocket plumes produce brief, intense episodes with the highest $v_{\rm ejecta}$ (Sec.~\ref{sec:mobilization-transport}, Fig.~\ref{fig:stream}).
\item \textit{Regional priors that matter.} Fe/Ti, maturity, packing, and $T$ differ systematically across maria, highlands, swirls, pyroclastics, and PSRs, shifting $k$, $\varepsilon'$, $\tan\delta$, $(\phi,c)$, and charging behavior; Tables~\ref{tab:regional} and \ref{tab:properties} provide design ranges and recommended reference values (Secs.~\ref{sec:reg_refvalues}, \ref{sec:eng_refvalues}).
\item \textit{Measurement priorities.} Co--register $E$--fields, grain charge, PSD, and flux at illumination/plasma boundaries; time--resolve dust flux during meteoroid streams; centimeter--scale thermophysics in PSRs; and standardized plume--surface monitors with stereo imaging and ejecta counters.
\end{enumerate}

The practical impact of this synthesis is threefold. First, it provides region--specific reference values that span maria, highlands, swirls, pyroclastics, and PSRs, based on Apollo through present missions. Second, it supplies bridging relations that propagate grain--scale structure into $k(T,\rho)$, $\varepsilon'$, and lofting thresholds including adhesion. Third, it identifies colocated measurements that most efficiently retire risk: $E$--field and charge with PSD at illumination boundaries; time--resolved dust flux during meteoroid streams; centimeter--scale thermophysics in PSRs; and standardized plume--surface monitors with stereo imaging and ejecta counters on CLPS and Artemis landers.

\section*{Acknowledgments} 
The work described here was carried out at the Jet Propulsion Laboratory, California Institute of Technology, Pasadena, California, under a contract with the National Aeronautics and Space Administration.
% \textcopyright 2025. California Institute of Technology. Government sponsorship acknowledged.

%\bibliography{lunar-dust}

\end{document}